\documentclass[twocolumn]{aastex631}

\usepackage{amsmath, physics, here}

\newcommand{\Add}[1]{\textcolor{black}{#1}}
\newcommand{\xHI}{x_{\rm H\,\textsc{i}}}
\newcommand{\revise}[1]{\textcolor{black}{#1}}

\shorttitle{Ly$\alpha$ Emission at $z=5-14$ and Cosmic Reionization}
\shortauthors{Kageura et al.}

\begin{document}

\title{Census of Ly$\alpha$ Emission from $\sim 600$ Galaxies at $z=5-14$:\\
Evolution of the Ly$\alpha$ Luminosity Function and a Late Sharp Cosmic Reionization}

\author[0009-0004-0381-7216]{Yuta Kageura}
\affiliation{Institute for Cosmic Ray Research, The University of Tokyo, 5-1-5 Kashiwanoha, Kashiwa, Chiba 277-8582, Japan}
\affiliation{Department of Physics, Graduate School of Science, The University of Tokyo, 7-3-1 Hongo, Bunkyo, Tokyo 113-0033, Japan}

\author[0000-0002-1049-6658]{Masami Ouchi}
\affiliation{National Astronomical Observatory of Japan, 2-21-1 Osawa, Mitaka, Tokyo 181-8588, Japan}
\affiliation{Institute for Cosmic Ray Research, The University of Tokyo, 5-1-5 Kashiwanoha, Kashiwa, Chiba 277-8582, Japan}
\affiliation{Department of Astronomical Science, SOKENDAI (The Graduate University for Advanced Studies), Osawa 2-21-1, Mitaka, Tokyo, 181-8588, Japan}
\affiliation{Kavli Institute for the Physics and Mathematics of the Universe (WPI), University of Tokyo, Kashiwa, Chiba 277-8583, Japan}

\author[0009-0000-1999-5472]{Minami Nakane}
\affiliation{Institute for Cosmic Ray Research, The University of Tokyo, 5-1-5 Kashiwanoha, Kashiwa, Chiba 277-8582, Japan}
\affiliation{Department of Physics, Graduate School of Science, The University of Tokyo, 7-3-1 Hongo, Bunkyo, Tokyo 113-0033, Japan}

\author[0009-0008-0167-5129]{Hiroya Umeda}
\affiliation{Institute for Cosmic Ray Research, The University of Tokyo, 5-1-5 Kashiwanoha, Kashiwa, Chiba 277-8582, Japan}
\affiliation{Department of Physics, Graduate School of Science, The University of Tokyo, 7-3-1 Hongo, Bunkyo, Tokyo 113-0033, Japan}

\author[0000-0002-6047-430X]{Yuichi Harikane}
\affiliation{Institute for Cosmic Ray Research, The University of Tokyo, 5-1-5 Kashiwanoha, Kashiwa, Chiba 277-8582, Japan}

\author[0000-0003-0581-5973]{Shintaro Yoshiura}
\affiliation{National Astronomical Observatory of Japan, 2-21-1 Osawa, Mitaka, Tokyo 181-8588, Japan}

\author[0000-0003-2965-5070]{Kimihiko Nakajima}
\affiliation{National Astronomical Observatory of Japan, 2-21-1 Osawa, Mitaka, Tokyo 181-8588, Japan}

\author[0000-0002-1319-3433]{Hidenobu Yajima}
\affiliation{Center for Computational Sciences, University of Tsukuba, Ten-nodai, 1-1-1 Tsukuba, Ibaraki 305-8577, Japan}

\author[0000-0002-8408-4816]{Tran Thi Thai}
\affiliation{National Astronomical Observatory of Japan, 2-21-1 Osawa, Mitaka, Tokyo 181-8588, Japan}

\correspondingauthor{Yuta Kageura}
\email{kageura@icrr.u-tokyo.ac.jp}



\begin{abstract}
We present the statistical properties of Ly$\alpha$ emission in 586 galaxies at $z=4.5-14.2$, observed by multiple JWST/NIRSpec spectroscopy projects, including JADES, GLASS, CEERS, and GO/DDT programs. We obtain Ly$\alpha$ equivalent width (EW), Ly$\alpha$ escape fraction, and ionizing photon production efficiency measurements or upper limits for these galaxies, and confirm that the Ly$\alpha$ emitting galaxy fraction decreases towards higher redshifts. We derive Ly$\alpha$ luminosity functions from $z\sim 5$ to $z\sim 10-14$ with the observed Ly$\alpha$ EW distributions and galaxy UV luminosity functions, and find a $\sim3$ dex decrease in number density at $L_\mathrm{Ly\alpha}=10^{42}-10^{43}$ erg s$^{-1}$ over the redshift range. \revise{Notably, this study presents the first constraints on the Ly$\alpha$ luminosity function at $z\sim 8-14$.} We obtain the neutral hydrogen fractions of $\xHI=0.17_{-0.16}^{+0.23}$, $0.63_{-0.28}^{+0.18}$, $0.79_{-0.21}^{+0.13}$, and $0.88_{-0.13}^{+0.11}$ at $z\sim6$, $7$, $8-9$, and $10-14$, respectively, via comparisons of the reionization models developed by semi-numerical simulations with 21cmFAST explaining the observations of Ly$\alpha$, UV continuum, and Planck electron optical depth. The high $\xHI$ values over $z\sim 7-14$ suggest a late and sharp reionization, with the primary reionization process occurring at $z\sim 6-7$. Such a late and sharp reionization is not easily explained by either a clumpy inter-galactic medium or sources of reionization in a classical faint-galaxy or a bright-galaxy/AGN scenario, unless a very high escape fraction or AGN duty cycle is assumed at $z\sim 6-7$. 
\end{abstract}

\keywords{Galaxy evolution (594), High-redshift galaxies (734), Lyman-alpha galaxies (978), Reionization (1383)}


\section{Introduction}
Cosmic reionization marks the last major phase transition in cosmic history.
During the Epoch of Reionization (EoR), neutral hydrogen in the intergalactic medium (IGM), formed at Recombination, was ionized by ultraviolet (UV) photons emitted by the first sources.
However, the exact progression of reionization with redshift (i.e., the reionization history) remains debated.
Understanding this history is crucial, as it provides insights into the primary sources driving reionization.\par
To investigate the reionization history, the redshift evolution of the neutral hydrogen fraction $\xHI$ has been estimated through various methods.
Gunn-Peterson troughs of quasar spectra \citep{gunn65} \Add{have shown} that reionization is largely completed (i.e., $\xHI\sim0$) by $z\sim6$ (e.g., \citealt{becker01, fan06}).
Recent advanced analyses of Ly$\alpha$ forests and Ly$\alpha+\beta$ dark fraction or gaps, and state-of-the-art simulations \Add{have suggested} that neutral hydrogen persists as late as $z\sim5.3$ \Add{(e.g., \citealt{kulkarni19, bosman22, spina24, zhu24})}.
At $z\sim6-7$, $\xHI$ has been investigated using various techniques, including Ly$\alpha$ damping wing absorption measurements in gamma-ray burst (GRB) and quasar (QSO) spectra (e.g., \citealt{totani06, totani14, wang20, yang20, durovcikova24, fausey24}), Ly$\alpha$ equivalent width (EW) distributions (e.g., \citealt{mason18, hoag19, bolan22}), Ly$\alpha$ luminosity functions (e.g., \citealt{ouchi10, konno14, thai23, umeda24b}), and clustering analyses of Ly$\alpha$ emitters (LAEs; e.g., \citealt{sobacchi15, ouchi18, umeda24b}).
The optical depth of the cosmic microwave background (CMB; \citealt{planck20}) also provides critical constraints on the reionization history.\par
The launch of the James Webb Space Telescope (JWST) has enabled investigations of the reionization history at even higher redshifts, $z\sim8-13$, through measurements of Ly$\alpha$ damping wing absorption in Lyman-break galaxies (LBGs; e.g., \citealt{curtislake23, hsiao24, umeda24a}) and Ly$\alpha$ EW distributions (e.g., \citealt{nakane24, tang24b, jones24}).
While \Add{the} numerous constraints on $\xHI$, including measurements at high redshifts, have been obtained, many suffer from low accuracy due to limited sample sizes, or exhibit inconsistencies across \Add{the} studies.
Thus, it is crucial to investigate the reionization history comprehensively, from its onset (i.e., high redshifts) to the end (i.e., low redshifts), using a consistent methodology and \Add{large statistical sample}.
Among various methods, the Ly$\alpha$ EW distribution is a powerful probe of the reionization history.
\revise{Constraints on $\xHI$ are obtained by examining the redshift evolution of the Ly$\alpha$ EW distribution from $z\sim5-6$ to higher redshifts.}
Although \citet{nakane24}, \citet{tang24b}, and \citet{jones24} explore this approach, their galaxy sample sizes are limited ($\sim50-200$ galaxies), \revise{and the $z\sim5-6$ baseline distribution is constructed using ground-based spectra, which require slit loss correction to compare with JWST/NIRSpec observations.}\par
In this study, we present the statistical properties of Ly$\alpha$ emission in \Add{586} galaxies at $z=4.5-14.2$ observed with JWST/NIRSpec.
\revise{Unlike previous studies, we establish a consistent Ly$\alpha$ EW baseline distribution directly from JWST/NIRSpec observations at $z\sim5$.
Using this improved baseline,} 
we investigate the evolution of EW distributions to estimate $\xHI$ at $z\sim6-14$.
The resulting $\xHI$ values provide insights into the reionization history and the primary ionizing sources.
\revise{Thanks to the large sample size, we also derive the Ly$\alpha$ luminosity function and place the first constraints on the Ly$\alpha$ luminosity function at $z\sim8-14$.}\par
This paper is organized as follows.
Section \ref{sec;data} describes the JWST/NIRSpec spectroscopic data and our sample selection.
Section \ref{sec;measure} presents the measurements of Ly$\alpha$ equivalent widths, Ly$\alpha$ escape fractions, and Ly$\alpha$ luminosity functions.
In Section \ref{sec;model}, we describe the \Add{models of} EW distributions derived from semi-numerical simulations.
Section \ref{sec;crh} details the estimated $\xHI$ values.
The implications for the reionization history and ionizing sources are discussed in Section \ref{sec;discussion}.
Section \ref{sec;summary} summarizes our findings.
Throughout this paper, we adopt the Planck cosmological parameters from the TT, TE, EE+lowE+lensing+BAO results \citep{planck20}: $\Omega_m=0.31$, $\Omega_\Lambda=0.69$, $\Omega_b=0.049$, $h=0.68$, $Y_p=0.24$, and $\tau=0.0561\pm0.0071$.
All magnitudes are in the AB system \citep{oke83}.

\section{Data and Sample}\label{sec;data}
We use JWST/NIRSpec micro-shutter assembly (MSA) spectra obtained from the following public observations: the JWST Advanced Deep Extragalactic Survey (JADES; GTO 1180/1181, GTO 1210/1286, and GO 3215, led by PIs D. Eisenstein, N. Lützgendorf, and D. Eisenstein \& R. Maiolino, respectively; \citealt{bunker23, eisenstein23a, eisenstein23b, deugenio24}), the GLASS JWST Early Release Science Program (ERS 1324, PI: T. Treu; \citealt{treu22}), the Cosmic Evolution Early Release Science (CEERS; ERS 1345, PI: S. Finkelstein; \citealt{arrabalharo23a, finkelstein23}), GO 1433 (PI: D. Coe; \citealt{hsiao24}) and DDT 2750 (PI: P. Arrabal Haro; \citealt{arrabalharo23b}).
All JADES programs, except GO 3215, observed \Add{the targets} with the prism ($R\sim100$) that covered $0.6\mathrm{-}5.3~\mathrm{\mu m}$ and three medium resolution ($R\sim1000$) grating/filter pairs of G140M/F070LP, G235M/F170LP, and G395M/F290LP that covered $0.7\mathrm{-}1.3~\mathrm{\mu m}$, $1.7\mathrm{-}3.1~\mathrm{\mu m}$, and $2.9\mathrm{-}5.1~\mathrm{\mu m}$, respectively.
These JADES programs, GTO 1180, 1181, 1210, and 1286, took the MSA spectra at the positions of 9, 9, 1, and 1 pointings, respectively.
The exposure times per pointing were $1.1\mathrm{-}27.8$ hours for the prism and $0.9\mathrm{-}6.9$ hours for each grating.
The JADES GO 3215 program did not use G235M/F170LP, and observed \Add{the targets} at one pointing position with the prism, G140M/F070LP, and G395/F290LP whose exposure times were 46.7, 11.7, and 46.7 hours, respectively.
The GLASS \Add{data were taken} at one pointing position with high resolution ($R\sim2700$) grating/filter pairs of G140H/F100LP ($1.0\mathrm{-}1.8~\mathrm{\mu m}$), G235H/F170LP ($1.7\mathrm{-}3.1~\mathrm{\mu m}$), and G395M/F290LP ($2.9\mathrm{-}5.1~\mathrm{\mu m}$).
For each grating/filter pair, the exposure time was 4.9 hours. 
The CEERS observed \Add{the targets} at eight and six pointing positions with the prism and three medium resolution grating/filter pairs, respectively, which are G140M/F100LP ($1.0\mathrm{-}1.8~\mathrm{\mu m}$), G235M/F170LP, and G395M/F290LP.
For each prism or grating per one pointing, the exposure time was 0.9-1.7 hours.
The GO 1433 and DDT 2750 programs took the prism data with integration times of 3.7 and 5.1 hours, respectively.\par
The JADES team has reduced the spectroscopic data using the pipeline developed by the ESA NIRSpec SOT and GTO NIRSpec teams, making it publicly available in Data Release 1 and 3\footnote{\url{https://jades-survey.github.io/scientists/data.html}}.
Since the JADES data include spectra without reliable redshift determinations, we select galaxies whose spectroscopic redshifts are determined using emission lines and/or a strong continuum break (flag A, B, and C in the JADES catalogs; see \citealt{deugenio24}).
In addition to the JADES galaxies in Data Release 3, we include three recently identified galaxies at $z>13$ in the JADES program: JADES-GS-z14-0, JADES-GS-z14-1 \citep{carniani24a, carniani24b}, and JADES-GS-z13-1-LA \citep{witstok24}.
\citet{nakajima23} and \citet{harikane24} have reduced the GLASS, CEERS, GO 1433, and DDT 2750 data using the JWST pipeline version 1.8.5 with the Calibration Reference Data System context file of \texttt{jwst\_1028.pmap} or \texttt{jwst\_1027.pmap}, with additional processes improving the flux calibration, noise estimate, and composition.
They produced catalogs containing only galaxies with redshifts determined via emission lines and/or a strong continuum break.
The catalogs of the JADES team and \citet{nakajima23} \& \citet{harikane24} contain 2562 galaxies with redshifts determined at $0.1\leq z_\mathrm{spec}\leq14.2$ in total.
To investigate the reionization history using Ly$\alpha$ properties, we select galaxies at $z>4.5$ with grating or prism spectra that cover the rest-frame Ly$\alpha$ line at $\lambda_a=1215.67$~\AA.
This selection yields \Add{586} galaxies: \Add{456} from JADES, 6 from GLASS, \Add{117} from CEERS, 1 from GO 1433, and \Add{6} from DDT 2750.
Among these, \Add{398} galaxies have grating spectra that cover the Ly$\alpha$ wavelength.

\section{Measurements of Lyman-alpha Equivalent Widths}\label{sec;measure}
\subsection{Lyman-alpha Flux Measurements}\label{flux_measure}
\begin{figure}[t]
    \centering
    \includegraphics[width=\linewidth]{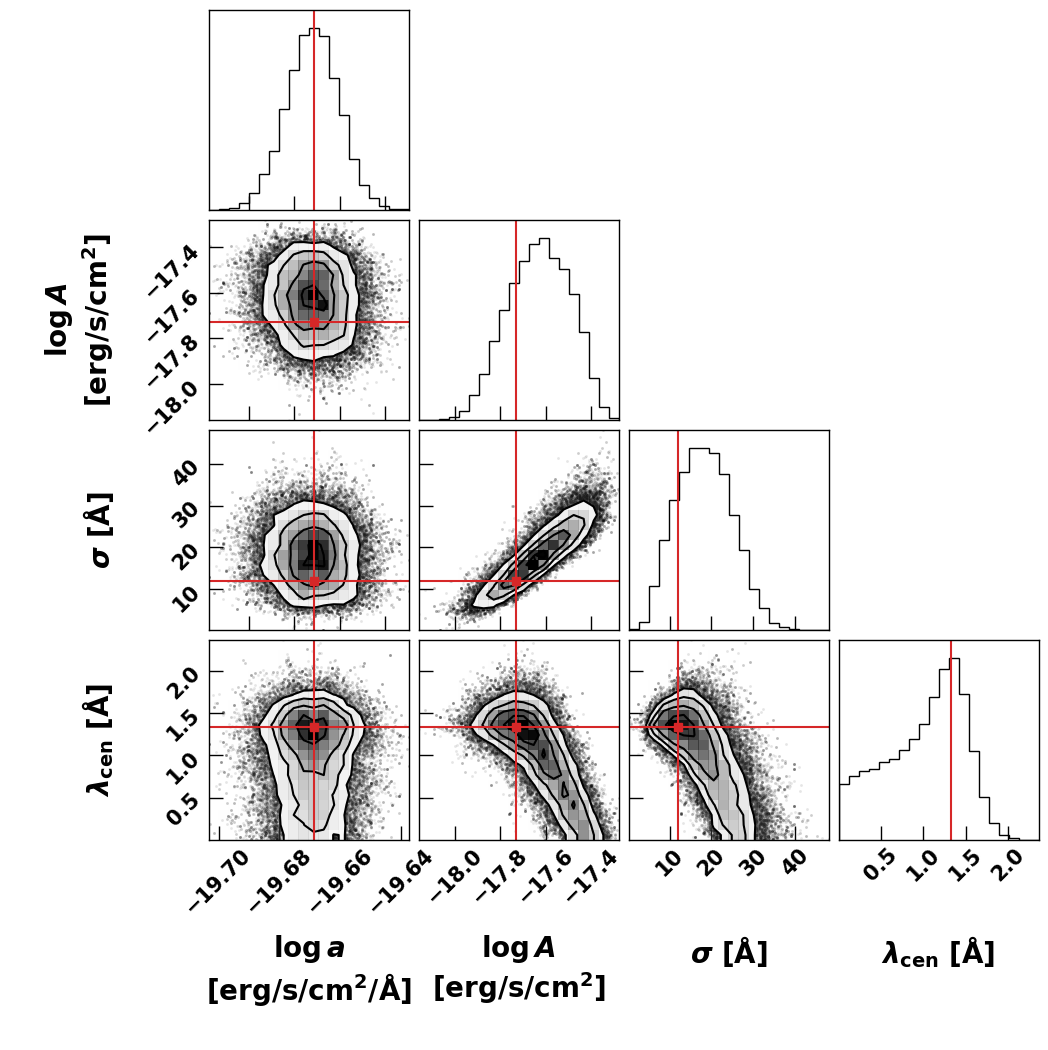}
    \caption{Posterior distributions of the fitting parameters for the grating spectrum of \Add{JADES-5591 (GN-z11)}, obtained by MCMC simulations. The red lines indicate the best-fit values for each parameter.}
    \label{mcmc}
\end{figure}
\begin{figure}[t]
    \centering
    \includegraphics[width=\linewidth]{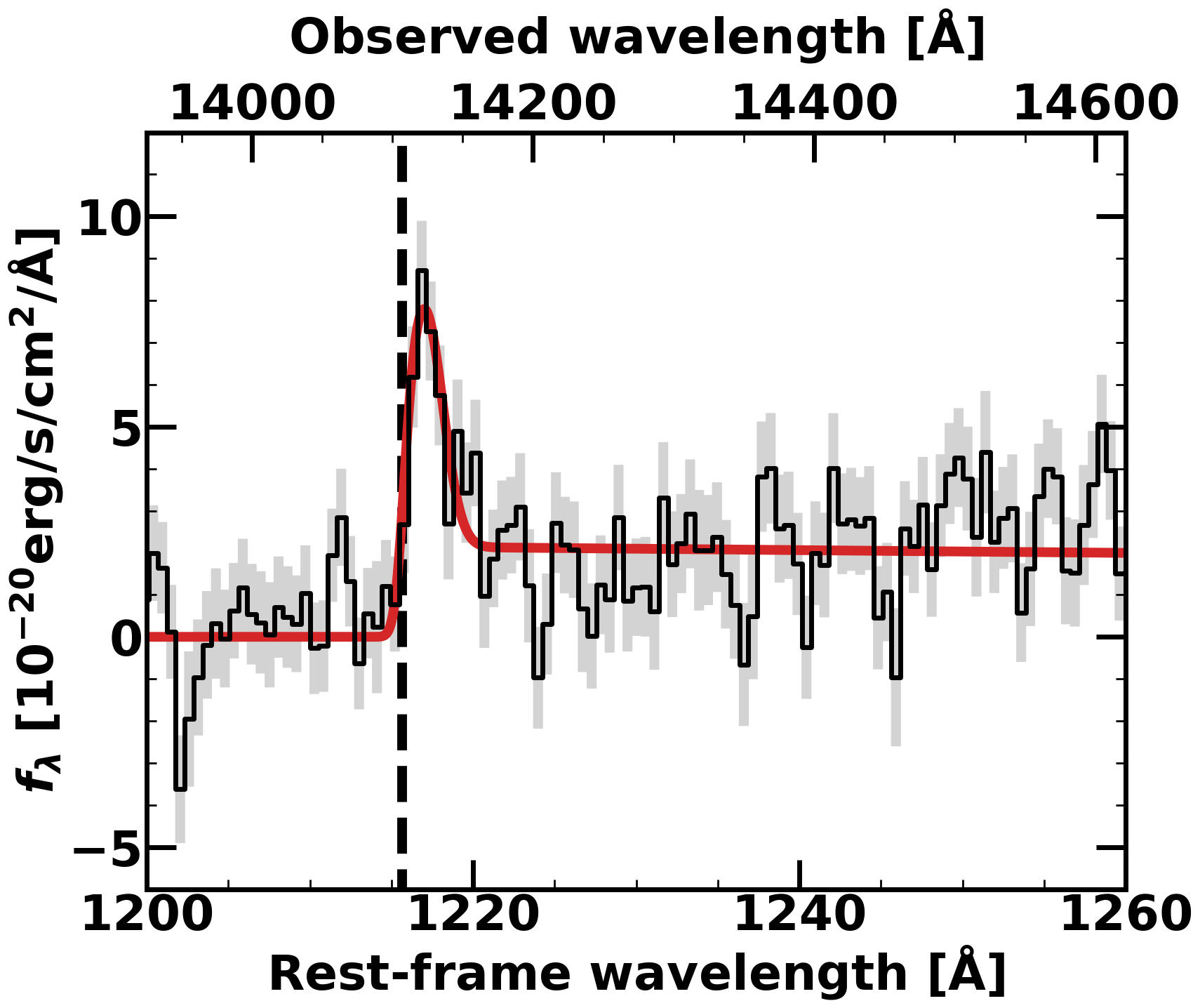}
    \caption{Grating spectrum of \Add{JADES-5591 (GN-z11)} around the Ly$\alpha$ emission line. The red solid line represents the best-fit model spectrum obtained from MCMC simulations. The black solid line and shaded region show the observed spectrum and associated $1\sigma$ error, respectively. The black dashed line marks the Ly$\alpha$ wavelength.}
    \label{fitting}
\end{figure}
By fitting the observed spectra, we derive the Ly$\alpha$ emission properties of galaxies, including the EW.
The intrinsic Ly$\alpha$ emission line is modeled as a Gaussian, while the continuum is assumed to follow a power-law profile.
The intrinsic spectrum is expressed as:
\begin{align}\label{f_lambda}
    f_\lambda=&a\qty(\frac{\lambda}{\lambda_a(1+z)})^{\beta}\notag\\
    &+\frac{A}{\sqrt{2\pi}\sigma}\exp\qty(-\frac{(\lambda-(\lambda_a+\Delta\lambda)(1+z))^2}{2\sigma^2}),
\end{align}
where $a$, $\beta$, $A$, $\sigma$, and $\Delta\lambda$ are the fitting parameters\revise{, corresponding to the continuum normalization, UV slope, Ly$\alpha$ flux, velocity dispersion, and wavelength shift, respectively.}
We assume that the IGM completely absorbs photons with wavelengths shorter than $\lambda=\lambda_a(1+z)$.
To account for instrumental effects, we convolve the intrinsic spectrum with the line spread function (LSF) of the NIRSpec instrument, as evaluated by \citet{isobe23} using calibration data of a planetary nebula.
Approximating the LSF as a Gaussian with standard deviation $\sigma_0$, the observed spectrum is expressed as:
\begin{align}
    f_\lambda=&\frac{a}{2}\qty(\frac{\lambda}{\lambda_a(1+z)})^{\Add{\beta}}\mathrm{erfc}\qty(\frac{\lambda_a(1+z)-\lambda}{\sqrt{2}\sigma_0})\notag\\
    &+\frac{A}{2\sqrt{2\pi}\sqrt{\sigma^2+\sigma_0^2}}\exp\qty(-\frac{(\lambda-(\lambda_a+\Delta\lambda)(1+z))^2}{2(\sigma^2+\sigma_0^2)})\notag\\
    &\times\mathrm{erfc}\qty(\frac{\sigma^2\lambda_a(1+z)-\sigma^2\lambda-\sigma_0^2\Delta\lambda(1+z)}{\sqrt{2}\sigma\sigma_0\sqrt{\sigma^2+\sigma_0^2}}).
\end{align}
\par
We derive the fitting parameters using \texttt{emcee} \citep{foremanmackey13} to perform Markov Chain Monte Carlo (MCMC) simulations.
\Add{The wavelength range of 1200~\AA-1800~\AA\ is used for the fitting procedure.}
\revise{We use the flat priors of
\begin{align}
    -21&<\log{a}~\mathrm{[erg~s^{-1}~cm^{-2}~\AA^{-1}]}<-18,\notag\\
    -21&<\log{A}~\mathrm{[erg~s^{-1}~cm^{-2}]}<-16,\notag\\
    0.1&<\sigma~\mathrm{[\AA]}<50,\notag\\
    0&<\Delta\lambda~\mathrm{[\AA]}<5.
\end{align}}
For galaxies with available grating spectra, we use these spectra for the fitting process.
Due to the low signal-to-noise ratio in the grating spectra, it is challenging to simultaneously fit the continuum amplitude $a$ and slope $\beta$.
Therefore, we fix $\beta$ to $-2$\Add{, which is typical for star-forming galaxies (e.g., \citealt{saxena24b, yanagisawa24})}.
For other galaxies, we use the prism spectra and fit all parameters, including $\beta$.
\Add{The Ly$\alpha$ properties derived from the grating and prism spectra are shown to be consistent by \citet{nakane24}.}
We determine the best-fit parameters and their $1\sigma$ uncertainties by the mode and $68\%$ highest posterior density interval (HPDI) of the posterior distribution.
Figure \ref{mcmc} and \ref{fitting} show an example of the MCMC posterior distributions and spectrum of JADES-5591 (GN-z11), respectively.
Figure \ref{tab:sample} and Table \ref{spec1} summarize the Ly$\alpha$ properties of all galaxies in our sample.

\subsection{Lyman-alpha Equivalent Width}\label{sec;ew}
\begin{figure*}[t]
    \centering
    \includegraphics[width=\linewidth]{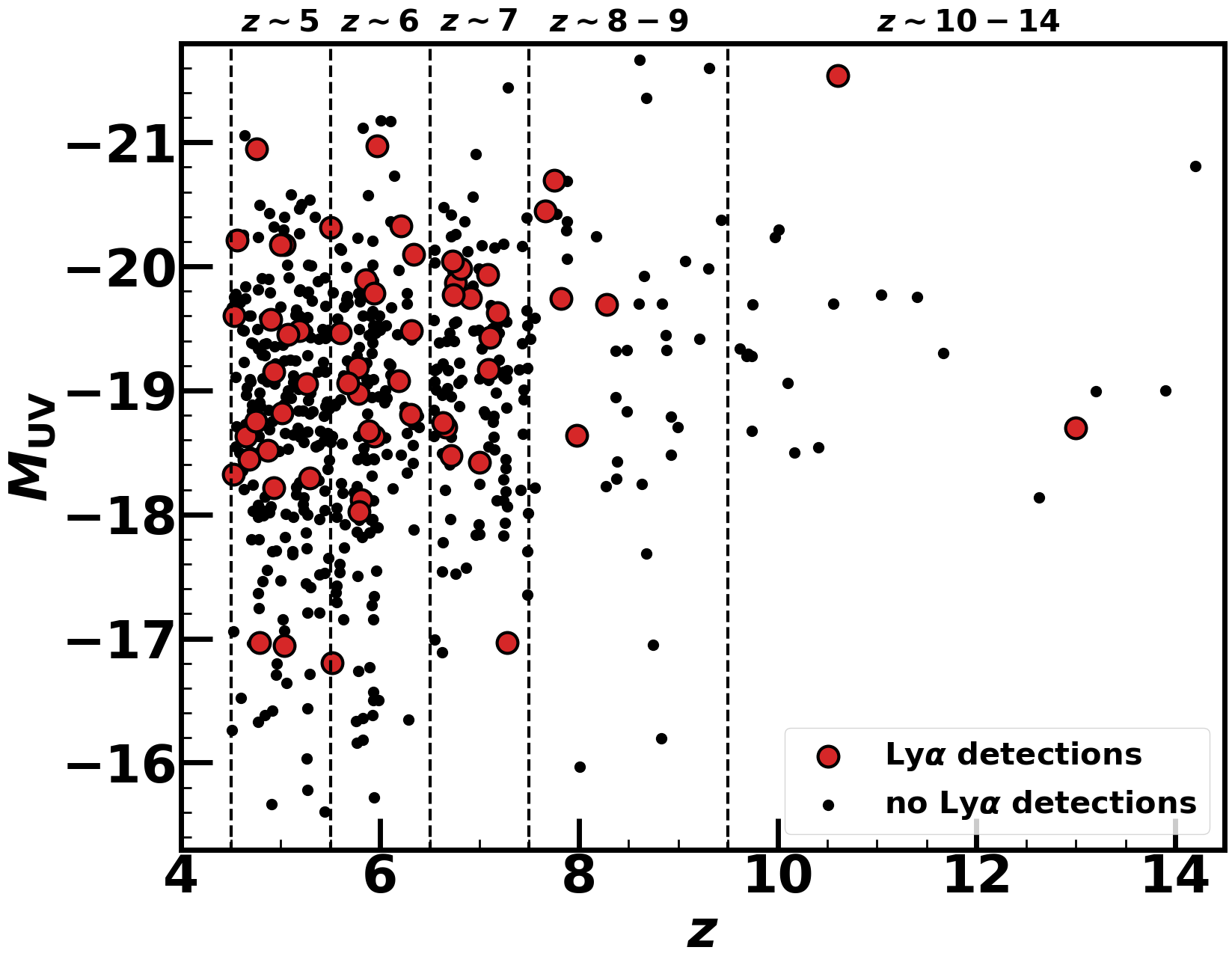}
    \caption{Distribution of absolute UV magnitudes $M_\mathrm{UV}$ and redshifts $z$ in our sample. The red (black) circles represent galaxies with (without) $3\sigma$ Ly$\alpha$ detections. Our sample is divided into five redshift bins for subsequent analyses: $4.5<z\leq5.5$, $5.5<z\leq6.5$, $6.5<z\leq7.5$, $7.5<z\leq9.5$, and $9.5<z\leq14.2$.}
    \label{MUVdist}
\end{figure*}
Using the fitting parameters derived in Section \ref{flux_measure}, we calculate the flux of the Ly$\alpha$ emission line as:
\begin{align}
    &F_\mathrm{Ly\alpha}\notag\\
    &=\int_{\lambda_a(1+z)}^\infty\dd{\lambda}\frac{A}{\sqrt{2\pi}\sigma}\exp\qty(-\frac{(\lambda-(\lambda_a+\Delta\lambda)(1+z))^2}{2\sigma^2}).
\end{align}
The rest-frame EW is then calculated as:
\begin{align}
    \mathrm{EW}=\frac{F_\mathrm{Ly\alpha}}{a(1+z)},
\end{align}
where $a$ is the continuum flux density at $\lambda=\lambda_a(1+z)$ in Equation (\ref{f_lambda}).
For galaxies without a Ly$\alpha$ detection at the $3\sigma$ significance level, we calculate the $3\sigma$ upper limit of the EW.
For JADES-GS-z14-0, JADES-GS-z14-1, and JADES-GS-z13-1-LA, we adopt the EW values reported by \citet{carniani24a} and \citet{witstok24}.
\Add{Among the 586 galaxies in our sample, Ly$\alpha$ emission lines are detected at the $3\sigma$ significance level for 60 galaxies.}\par
In Figure \ref{MUVdist}, we show the distribution of $M_\mathrm{UV}$ and redshift for galaxies in our sample, with and without Ly$\alpha$ detection.
\Add{We derive the $M_\mathrm{UV}$ values} by integrating the prism spectra over the rest-frame wavelength range of 1400~\AA\ to 1600~\AA.
For galaxies lacking prism spectra, the grating spectra are used instead.
\Add{For galaxies observed in the GLASS and GO 1433 programs, we use the $M_\mathrm{UV}$ values obtained by \citet{nakajima23} and \citet{harikane24}, which account for lensing effects.}
\revise{Sections \ref{sec;lya_frac}, \ref{sec;lya_fesc}, and \ref{sec;lyaLF} provide a comprehensive summary of the key aspects of our measurements.}

\subsection{Lyman-alpha Fraction}\label{sec;lya_frac}
\begin{figure*}[t]
    \centering
    \includegraphics[width=\linewidth]{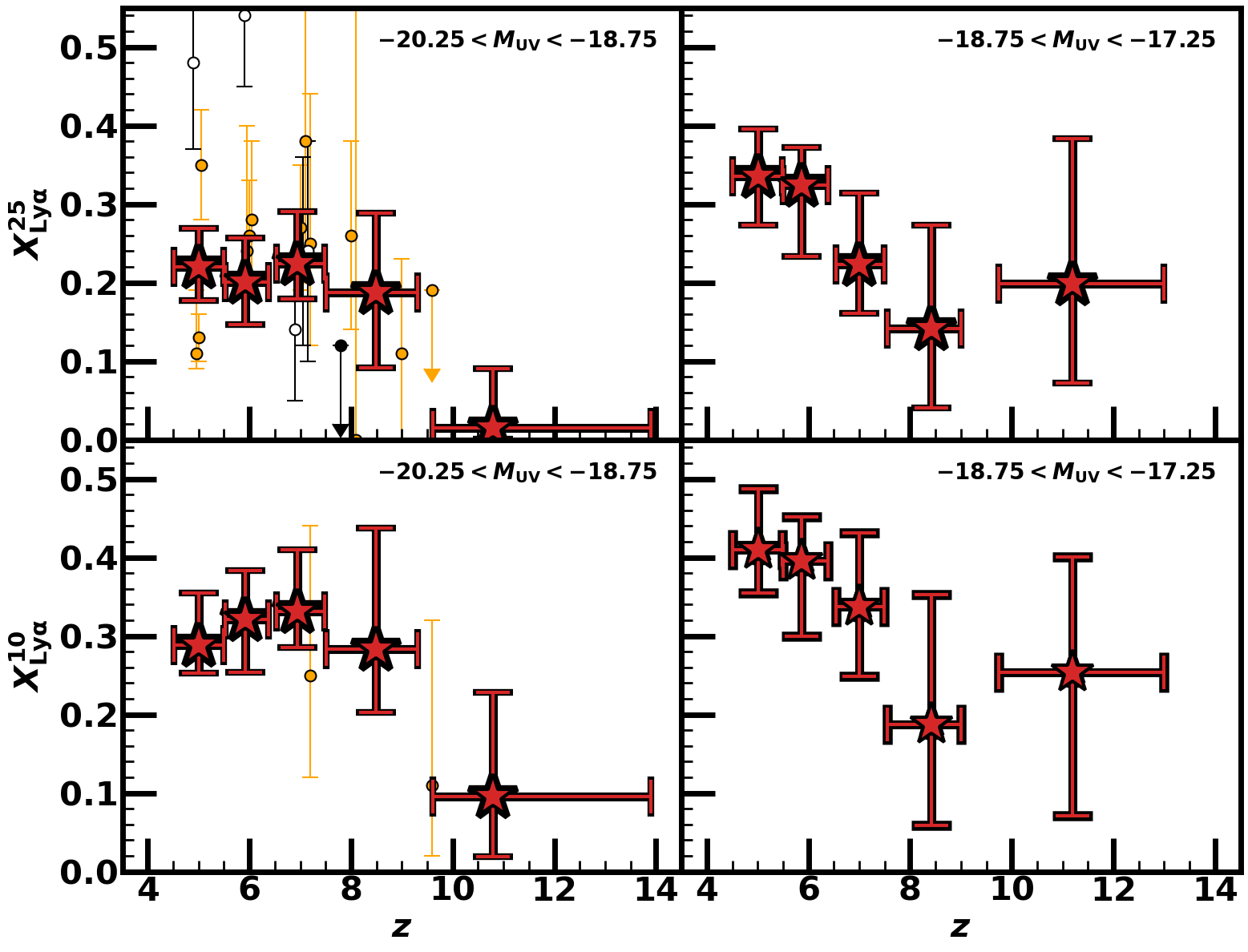}
    \caption{Ly$\alpha$ fraction as a function of redshift. The red stars represent $X_\mathrm{Ly\alpha}^{\mathrm{EW_{th}}}$, the fraction of galaxies in our sample with Ly$\alpha$ EW exceeding $\mathrm{EW_{th}}$. The top two panels show $X_\mathrm{Ly\alpha}^{25}$, while the bottom two panels display $X_\mathrm{Ly\alpha}^{10}$. The left (right) two panels correspond to galaxies with $-20.25<M_\mathrm{UV}<-18.75$ ($-18.75<M_\mathrm{UV}<-17.25$). The white circles indicate pre-JWST measurements of $X_\mathrm{Ly\alpha}^{25}$ from the literature (\citealt{stark11, ono12, schenker12, pentericci18, mason19}), while the orange circles represent recent JWST results (\citealt{jones24, nakane24, napolitano24, tang24a}). Across all panels, $X_\mathrm{Ly\alpha}^\mathrm{EW_{th}}$  decreases with redshift, suggesting increasing damping wing absorption by neutral hydrogen in the IGM.}
    \label{lyafraction}
\end{figure*}
In this section, we derive \Add{the Ly$\alpha$ fraction $X_\mathrm{Ly\alpha}^\mathrm{EW_{th}}$, defined as the fraction of galaxies with Ly$\alpha$ EW values exceeding the threshold $\mathrm{EW_{th}}$, in our sample.}
We define $n_\mathrm{Ly\alpha}$ as the number of galaxies with $\mathrm{EW}>\mathrm{EW_{th}}$ and $N$ as the total number of galaxies in the sample.
Since many galaxies in our sample have large EW upper limits, the true value of $n_\mathrm{Ly\alpha}$ cannot be directly equated to the number of galaxies with Ly$\alpha$ detections at the $3\sigma$ level and the best-fit value $\mathrm{EW}>\mathrm{EW_{th}}$.
\revise{The Kaplan-Meier estimator is generally an effective method for statistically rigorous treatment of non-detections \citep{kaplan58}. However, applying the Kaplan-Meier estimator in this study is challenging due to the difficulty in defining a clear classification between detections and non-detections. Although we set the threshold at $3\sigma$ in Section \ref{sec;ew}, this choice is inherently user-defined and may influence the fraction of classified non-detections.}
To avoid underestimating $n_\mathrm{Ly\alpha}$, we account for completeness in the analysis (see also \citealt{tang24a} and  \citealt{jones24} for completeness evaluations).
By considering the uncertainties in the EW measurements for individual galaxies, we perform Monte Carlo simulations to estimate $n_\mathrm{Ly\alpha}$.\par
\revise{We also account for the effect of binomial noise by using the statistics for small numbers of events by \citet{gehrels86}:
\begin{align}
    p(X_\mathrm{Ly\alpha}^\mathrm{EW_{th}}\mid n_\mathrm{Ly\alpha})=&\frac{1}{B(n_\mathrm{Ly\alpha}+1, N-n_\mathrm{Ly\alpha}+1)}\notag\\
    &\hspace{0.5em}\times(X_\mathrm{Ly\alpha}^\mathrm{EW_{th}})^{n_\mathrm{Ly\alpha}}(1-X_\mathrm{Ly\alpha}^\mathrm{EW_{th}})^{N-n_\mathrm{Ly\alpha}},
\end{align}
where $B(x,y)$ is the beta function.}
\par
In this way, we calculate $X_\mathrm{Ly\alpha}^\mathrm{EW_{th}}$ while accounting for the completeness effect and binomial noise.
Previous studies have investigated the Ly$\alpha$ fraction with $\mathrm{EW_{th}}=25$~\AA\ for galaxies with $-20.25<M_\mathrm{UV}<-18.75$ (e.g., \citealt{stark11, ono12, schenker12, pentericci18, mason19, jones24, nakane24, napolitano24, tang24a}).
In Figure \ref{lyafraction}, we compare our $X_\mathrm{Ly\alpha}^\mathrm{EW_{th}}$ measurements with those from the literature.
We adopt two UV magnitude bins, $-20.25<M_\mathrm{UV}<-18.75$ and $-18.75<M_\mathrm{UV}<-17.25$, as well as two threshold values, 25~\AA\ and 10~\AA.
Although our results at $z\sim5-6$ are lower than pre-JWST measurements, they are consistent with other JWST studies.
Across all four panels in Figure \ref{lyafraction}, $X_\mathrm{Ly\alpha}^\mathrm{EW_{th}}$ shows a decreasing trend with redshift, suggesting damping wing absorption by neutral hydrogen in the IGM.
For instance, $X_\mathrm{Ly\alpha}^{25}$ ($X_\mathrm{Ly\alpha}^{10}$) decreases from $22_{-4}^{+5}\%$ ($29_{-4}^{+7}\%$) at $z\sim5$ to $<10\%$ ($10_{-8}^{+13}\%$) at $z\sim11$ for galaxies with $-20.25<M_\mathrm{UV}<-18.75$.
The decreasing trend is particularly pronounced for galaxies with $-18.75<M_\mathrm{UV}<-17.25$, likely due to the smaller ionized bubble radius around faint galaxies, making Ly$\alpha$ photons more susceptible to absorption by the IGM.
\revise{We calculate Spearman’s rank correlation coefficient for the Ly$\alpha$ fraction in faint galaxies and find a strong negative correlation ($\rho = -0.90$) with a p-value of 0.037, indicating that the trend is statistically significant. We also fit the redshift evolution of the Ly$\alpha$ fraction with a linear function. The best-fit slopes are $-0.032 \pm 0.022$ for $X_\mathrm{Ly\alpha}^{25}$ and $-0.035 \pm 0.027$ for $X_\mathrm{Ly\alpha}^{10}$, suggesting a decreasing trend. However, the slopes remain consistent with zero at the $2\sigma$ level, indicating that the statistical significance is limited by the sample size.}
Notably, $X_\mathrm{Ly\alpha}^\mathrm{EW_{th}}$ at $z\sim11$ for $-18.75<M_\mathrm{UV}<-17.25$ galaxies is elevated because it includes the recently identified LAE at $z=13$, JADES-GS-z13-1-LA.
\Add{In Table \ref{table:xlya}, we summarize our Ly$\alpha$ fractions.}
\begin{deluxetable}{ccc|ccc}
    \tablecolumns{3}
    \tabletypesize{\scriptsize}
    \tablecaption{Ly$\alpha$ Fraction
    \label{table:xlya}}
    \tablehead{
    \multicolumn{3}{c|}{$-20.25<M_\mathrm{UV}<-18.75$} & \multicolumn{3}{c}{$-18.75<M_\mathrm{UV}<-17.25$} \\
    $z$ & $X_\mathrm{Ly\alpha}^{25}$ & $X_\mathrm{Ly\alpha}^{10}$ & $z$ & $X_\mathrm{Ly\alpha}^{25}$ & $X_\mathrm{Ly\alpha}^{10}$\\
    (1) & (2) & (3) & (1) & (2) & (3)
    }
    \startdata
    $5.01_{-0.50}^{+0.48}$ & $0.22_{-0.04}^{+0.05}$ & $0.29_{-0.04}^{+0.07}$ & $5.01_{-0.50}^{+0.47}$ & $0.34_{-0.06}^{+0.06}$ & $0.41_{-0.06}^{+0.08}$\\
    $5.92_{-0.40}^{+0.46}$ & $0.20_{-0.05}^{+0.06}$ & $0.32_{-0.07}^{+0.06}$ & $5.87_{-0.36}^{+0.52}$ & $0.32_{-0.09}^{+0.05}$ & $0.40_{-0.10}^{+0.06}$\\
    $6.95_{-0.41}^{+0.53}$ & $0.22_{-0.04}^{+0.07}$ & $0.33_{-0.05}^{+0.08}$ & $7.01_{-0.46}^{+0.48}$ & $0.22_{-0.06}^{+0.09}$ & $0.34_{-0.09}^{+0.09}$\\
    $8.49_{-0.98}^{+0.82}$ & $0.19_{-0.10}^{+0.10}$ & $0.28_{-0.08}^{+0.15}$ & $8.42_{-0.87}^{+0.58}$ & $0.14_{-0.10}^{+0.13}$ & $0.19_{-0.13}^{+0.17}$\\
    $10.80_{-1.18}^{+3.10}$ & $0.02_{-0.01}^{+0.08}$ & $0.10_{-0.08}^{+0.13}$ & $11.19_{-1.45}^{+1.81}$ & $0.20_{-0.13}^{+0.19}$ & $0.25_{-0.18}^{+0.15}$
    \enddata
    \tablecomments{(1): Mean redshift and the
lower/upper boundary of a subsample. (2): Fraction of galaxies with Ly$\alpha$ $\mathrm{EW}>25$~\AA. (3): Fraction of galaxies with Ly$\alpha$ $\mathrm{EW}>10$~\AA.}
\end{deluxetable}

\subsection{Lyman-alpha Escape Fraction}\label{sec;lya_fesc}
\begin{figure*}[t]
    \centering
    \includegraphics[width=\linewidth]{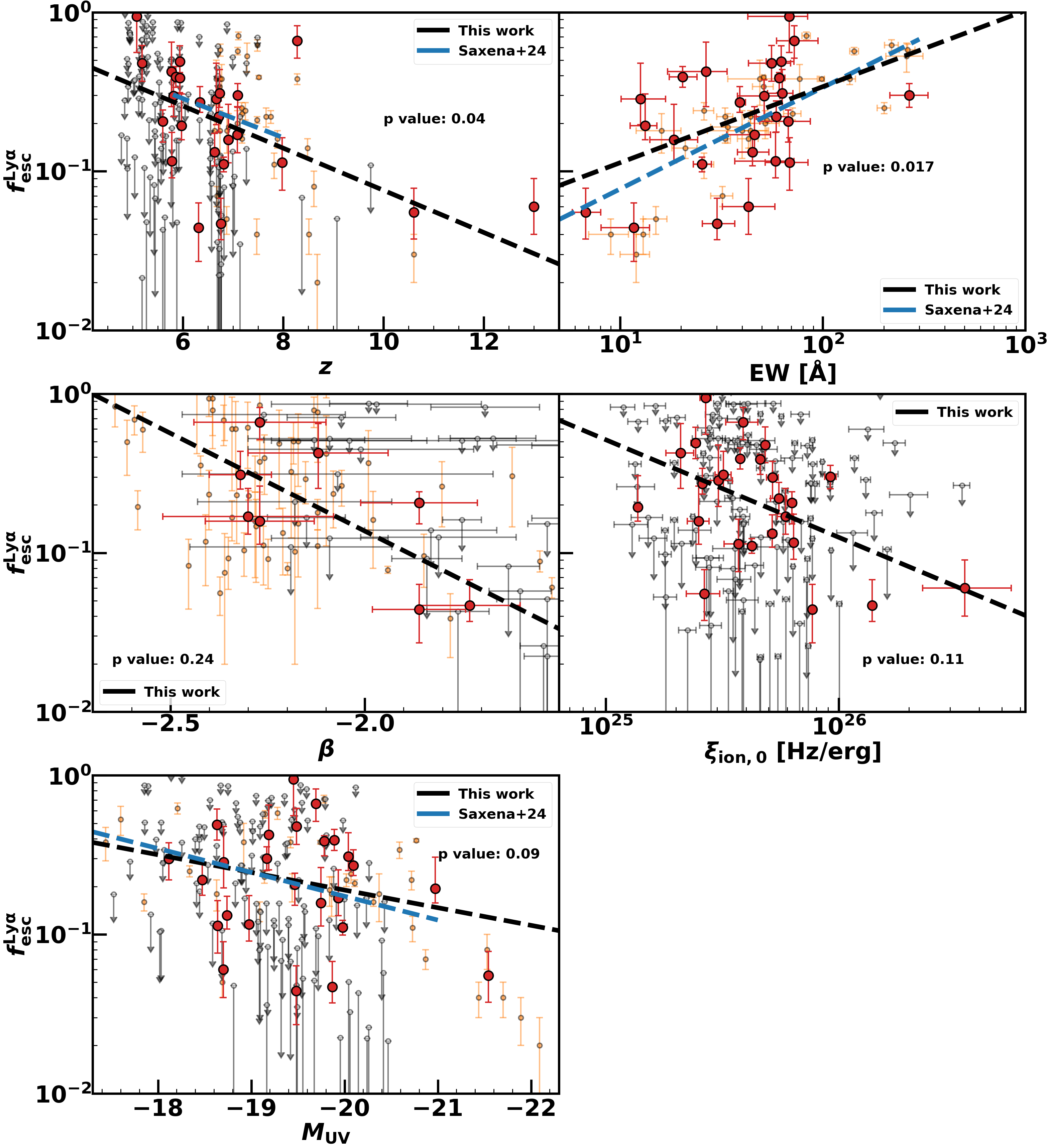}
    \caption{Escape fraction of Ly$\alpha$ photons ($f_\mathrm{esc}^{\mathrm{Ly\alpha}}$) as a function of various galaxy properties. Top left: $f_\mathrm{esc}^{\mathrm{Ly\alpha}}$ versus redshift ($z$). Top right: $f_\mathrm{esc}^{\mathrm{Ly\alpha}}$ versus Ly$\alpha$ EW. Middle left: $f_\mathrm{esc}^{\mathrm{Ly\alpha}}$ versus UV continuum slope ($\beta$). Middle right: $f_\mathrm{esc}^{\mathrm{Ly\alpha}}$ versus ionizing photon production efficiency under the assumption of zero LyC escape ($\xi_\mathrm{ion,0}$). Bottom left: $f_\mathrm{esc}^{\mathrm{Ly\alpha}}$ versus UV magnitude ($M_\mathrm{UV}$). The red circles represent galaxies with Ly$\alpha$ detection at the $3\sigma$ level, while the open circles indicate galaxies without Ly$\alpha$ detection. The yellow circles show measurements of \citet{tang24b} and \citet{tang24a}. \revise{The black dashed lines represent the correlations derived in this work, while the blue dashed lines indicate the relations reported by \citet{saxena24a}. The two sets of correlations show good agreement.}}
    \label{fesc_lya}
\end{figure*}
We calculate the Ly$\alpha$ escape fraction $f_\mathrm{esc}^{\mathrm{Ly\alpha}}$ for galaxies observed with the gratings in the JADES program in our sample by the equation:
\begin{align}
    f_\mathrm{esc}^{\mathrm{Ly\alpha}}=\frac{F_\mathrm{Ly\alpha,obs}}{F_\mathrm{Ly\alpha,int}},
\end{align}
where $F_\mathrm{Ly\alpha,obs}$ is the Ly$\alpha$ flux obtained in Section \ref{flux_measure} and $F_\mathrm{Ly\alpha,int}$ is the intrinsic Ly$\alpha$ flux.
We assume the electron temperature and density during the EoR are $T_e=1.5\times10^4~\mathrm{K}$ and $n_e=500~\mathrm{cm^{-3}}$, respectively \citep{isobe23, nakajima23}.
Using PyNeb \citep{luridiana15}, we derive the intrinsic flux ratios of Ly$\alpha$ and Balmer lines under the assumption of case B recombination: $F_\mathrm{Ly\alpha}/F_\mathrm{H_\alpha}=8.98$, $F_\mathrm{Ly\alpha}/F_\mathrm{H_\beta}=25.02$, and $F_\mathrm{Ly\alpha}/F_\mathrm{H_\gamma}=52.89$.
We use the flux values of Balmer lines in the JADES catalogs \citep{deugenio24} and calculate $f_\mathrm{esc}^{\mathrm{Ly\alpha}}$ for galaxies with H$\alpha$, H$\beta$, or H$\gamma$ flux measurements.
We assume the dust extinction of Balmer lines is negligible.
We also calculate the ionizing photon production efficiency under the assumption of zero LyC escape ($\xi_\mathrm{ion,0}$) from Balmer line fluxes.
Assuming ionizing photons do not escape from a galaxy, the amount of ionizing photons produced in a unit time is given by the equation \citep{osterbrock06}:
\begin{align}
    \dot{n}=7.28\times10^{11}~\mathrm{s^{-1}}~L(\mathrm{H_\alpha})/(\mathrm{erg\ s^{-1}}),
\end{align}
where $L(\mathrm{H_\alpha})$ is the H$\alpha$ luminosity.
For galaxies with H$\beta$ or H$\gamma$ measurements but without H$\alpha$ measurements, we use Balmer decrements $\mathrm{H\alpha/H\beta}=2.79$ and $\mathrm{H\alpha/H\gamma}=5.89$ to derive H$\alpha$ luminosity.
Then $\xi_\mathrm{ion,0}$ is obtained by the equation:
\begin{align}
    \xi_\mathrm{ion,0}=\frac{\dot{n}}{L_\mathrm{UV}},
\end{align}
where $L_\mathrm{UV}$ is the luminosity in units of $\mathrm{erg\ s^{-1}\ Hz^{-1}}$ at $\lambda=1500$~\AA.
\par
In Figure \ref{fesc_lya}, we show $f_\mathrm{esc}^{\mathrm{Ly\alpha}}$ as a function of redshift, Ly$\alpha$ EW, UV continuum slope ($\beta$), $\xi_\mathrm{ion,0}$, and $M_\mathrm{UV}$.
We use the $\beta$ values measured by \citet{yanagisawa24}.
For JADES-GS-z13-1-LA, we adopt the values of $f_\mathrm{esc}^\mathrm{Ly\alpha}$, $\beta$, $\xi_\mathrm{ion,0}$, and $M_\mathrm{UV}$ reported by \citet{witstok24}.
\revise{To quantify these relationships, we compute the correlations between the Ly$\alpha$ escape fraction and each property, and assess their statistical significance using Spearman’s rank correlation test.
The $f_\mathrm{esc}^{\mathrm{Ly\alpha}}$ values decrease with $z$ (p-value: 0.040) and increase with Ly$\alpha$ EW (p-value: 0.017), suggesting increasing damping wing absorption in the IGM.
Galaxies with a red UV slope (p-value: 0.24), high $\xi_\mathrm{ion,0}$ (p-value: 0.11), and bright $M_\mathrm{UV}$ (p-value: 0.090) tend to have a low $f_\mathrm{esc}^{\mathrm{Ly\alpha}}$, although statistically not significant.}
This correlation can be explained as follows:
In bright galaxies with a thick gaseous disc, the LyC escape fraction is low, and ionizing photons are absorbed within the interstellar medium (ISM), resulting in a high $\xi_\mathrm{ion,0}$.
The UV spectrum becomes redder due to an increased contribution from nebular emission.
Ly$\alpha$ photons are also less likely to escape in such galaxies \citep{choustikov24}, leading to a low $f_\mathrm{esc}^\mathrm{Ly\alpha}$.
\revise{The relationships between the Ly$\alpha$ escape fraction and $z$, Ly$\alpha$ EW, and $M_\mathrm{UV}$ obtained in this study are consistent with those reported by \citet{saxena24a}.}

\subsection{Lyman-alpha Luminosity Functions}\label{sec;lyaLF}
\begin{figure*}[t]
    \centering
    \includegraphics[width=\linewidth]{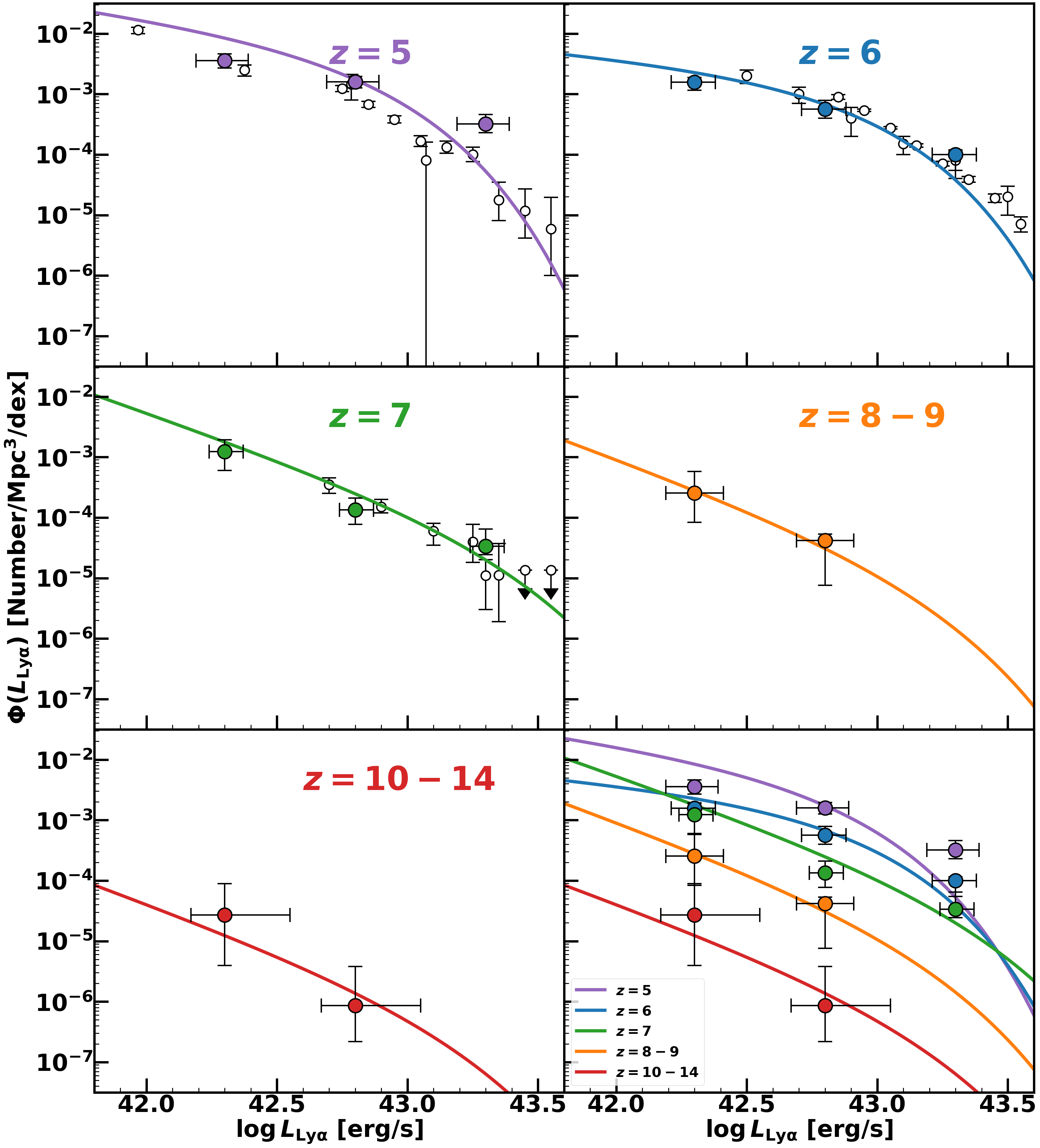}
    \caption{Evolution of Ly$\alpha$ luminosity function. The Ly$\alpha$ luminosity function at $z=5$ (top left), $z=6$ (top right), $z=7$ (middle left), $z=8-9$ (middle right), and $z=10-14$ (bottom left) are shown. The solid lines represent the best-fit Schechter functions. The open circles represent the measurements of Ly$\alpha$ luminosity function from \citet{ouchi08}, \citet{drake17}, \citet{ota17}, and \citet{umeda24b}. The bottom right panel summarizes the evolution of the Ly$\alpha$ luminosity function.}
    \label{lyaLF}
\end{figure*}
\begin{deluxetable}{c|ccc}
    \tablecolumns{4}
    \tablewidth{\linewidth}
    \tabletypesize{\scriptsize}
    \tablecaption{Ly$\alpha$ LF at $z=5-14$
    \label{table:lyalf_obs}}
    \tablehead{
    $z$ & \multicolumn{3}{c}{$\log{\Phi}_\mathrm{Ly\alpha}(L_\mathrm{Ly\alpha})~(\mathrm{Mpc^{-3}\ dex^{-1}})$}\\
     & $\log{L_\mathrm{Ly\alpha}}=42.3$ & $\log{L_\mathrm{Ly\alpha}}=42.8$ & $\log{L_\mathrm{Ly\alpha}}=43.3$\\
     & ($\mathrm{erg\ s^{-1}}$) & ($\mathrm{erg\ s^{-1}}$) & ($\mathrm{erg\ s^{-1}}$)
    }
    \startdata
    $z\sim5$ & $-2.45_{-0.12}^{+0.11}$ & $-2.80_{-0.10}^{+0.09}$ & $-3.49_{-0.15}^{+0.15}$\\
    $z\sim6$ & $-2.80_{-0.13}^{+0.08}$ & $-3.25_{-0.15}^{+0.14}$ & $-4.00_{-0.26}^{+0.06}$\\
    $z\sim7$ & $-2.91_{-0.31}^{+0.20}$ & $-3.87_{-0.24}^{+0.19}$ & $-4.47_{-0.14}^{+0.29}$\\
    $z\sim8-9$ & $-3.59_{-0.48}^{+0.36}$ & $-4.38_{-0.74}^{+0.11}$ & -\\
    $z\sim10-14$ & $-4.57_{-0.83}^{+0.52}$ & $-6.07_{-0.59}^{+0.65}$ & -
    \enddata
\end{deluxetable}
\begin{deluxetable}{cccc}
    \tablecolumns{4}
    \tablewidth{\linewidth}
    \tabletypesize{\scriptsize}
    \tablecaption{Best-fit Parameters for Ly$\alpha$ LF at $z=5-14$
    \label{table:lyalf_fit}}
    \tablehead{
    \colhead{$z$} & \colhead{$\log{\phi^\star}$} & \colhead{$\log{L^\star}$} & \colhead{$\alpha$}\\
     & ($\mathrm{Mpc^{-3}\ dex^{-1}}$) & ($\mathrm{erg\ s^{-1}}$)
    }
    \startdata
    $5.01_{-0.50}^{+0.47}$ & $-2.50_{-0.06}^{+0.06}$ & $42.69$ (fixed) & $-1.61$ (fixed)\\
    $5.90_{-0.40}^{+0.49}$ & $-3.03_{-0.13}^{+0.07}$ & $42.75$ (fixed) & $-1.39$ (fixed)\\
    $6.96_{-0.42}^{+0.52}$ & $-4.45_{-0.10}^{+0.09}$ & $43.23$ (fixed) & $-2.49$ (fixed)\\
    $8.41_{-0.90}^{+1.03}$ & $-4.98_{-0.35}^{+0.21}$ & $43.03$ (fixed) & $-2.56$ (fixed)\\
    $11.00_{-1.39}^{+3.18}$ & $-6.33_{-0.50}^{+0.44}$ & $43.03$ (fixed) & $-2.56$ (fixed)
    \enddata
\end{deluxetable}
In this section, we derive the Ly$\alpha$ luminosity functions from $z\sim5$ to $z\sim14$, using the Ly$\alpha$ EW distributions and UV luminosity functions.
The Ly$\alpha$ EW, Ly$\alpha$ luminosity $L_\mathrm{Ly\alpha}$, and UV magnitude $M_\mathrm{UV}$ are connected by the equation:
\begin{align}\label{EW_Llya}
    \mathrm{EW}=L_\mathrm{Ly\alpha}\frac{\lambda_a^2}{4\pi cd^2}10^{0.4(M_\mathrm{UV}+48.6)},
\end{align}
where $d=10~\mathrm{pc/cm}=3.09\times10^{19}$.
We assume a flat continuum at $\lambda=1216$~\AA\ - $1500$~\AA, which is the same assumption as the UV slope $\beta=-2$ for the grating spectra in Section \ref{flux_measure}.
\revise{Our sample may be biased when Ly$\alpha$ is detected through follow-up spectroscopy of continuum selected  galaxies. To account for this effect, the Ly$\alpha$ luminosity function is calculated as the convolution of the UF luminosity function and the EW distribution \citep{dijkstra12}}:
\begin{align}\label{lyaLF_eq}
    \Phi_\mathrm{Ly\alpha}(L_\mathrm{Ly\alpha})=\int\dd{M_\mathrm{UV}}\Phi_\mathrm{UV}(M_\mathrm{UV})\mathrm{EW}\ p(\mathrm{EW}, M_\mathrm{UV}),
\end{align}
where $\Phi_\mathrm{UV}(M_\mathrm{UV})$ is the UV luminosity function and $p(\mathrm{EW}, M_\mathrm{UV})$ is the Ly$\alpha$ EW distribution for galaxies with the UV magnitude $M_\mathrm{UV}$.
Equation (\ref{EW_Llya}) is substituted to Equation (\ref{lyaLF_eq}).\par
The Ly$\alpha$ EW distribution $p(\mathrm{EW}, M_\mathrm{UV})$ is derived from our measurements.
To account for the dependence of the EW distribution on $M_\mathrm{UV}$, We divide our sample into five $M_\mathrm{UV}$ bins: $M_\mathrm{UV}\leq-20.0$, $-20.0 < M_\mathrm{UV}\leq-19.0$, $-19.0< M_\mathrm{UV}\leq-18.0$, $-18.0<M_\mathrm{UV}\leq-17.0$, and $-17.0<M_\mathrm{UV}$.
\Add{We determine the $M_\mathrm{UV}$ bins so that each bin contains at least 20 galaxies at $z\sim5$.}
We numerically construct the distribution from the histogram of the observed EW values.
For the UV luminosity function $\Phi_\mathrm{UV}(M_\mathrm{UV})$, we adopt the form of the Schechter function:
\begin{align}
    \Phi_\mathrm{UV}(M_\mathrm{UV})=&\frac{\ln{10}}{2.5}\phi^\star10^{-0.4(M_\mathrm{UV}-M_\mathrm{UV}^\star)(\alpha+1)}\notag\\
    &\times\exp\qty(-10^{-0.4(M_\mathrm{UV}-M_\mathrm{UV}^\star)}).
\end{align}
The parameters $\phi^\star$, $M_\mathrm{UV}^\star$, and $\alpha$ are taken from the literature \citep{bouwens21, harikane24b}.
Note that the EW values measured in this study are not corrected for slit loss effects. From \citet{nakane24}, the EW of the entire galaxy, including the extended components outside the slit, is obtained by multiplying the measured EW value by two to correct for the slit loss effect of NIRSpec.\par
We show the results of Ly$\alpha$ luminosity functions for $\log{L_\mathrm{Lya}}=42.3$, $42.8$, and $43.3~\mathrm{erg\ s^{-1}}$ in Figure \ref{lyaLF} and Table \ref{table:lyalf_obs}.
The error of the Ly$\alpha$ luminosity function at each luminosity bin is estimated using bootstrap sampling of the Ly$\alpha$ EW distribution and the error of $\phi^\star$ in the UV luminosity functions presented in \citet{bouwens21} and \citet{harikane24b}.
The error in the x-direction in Figure \ref{lyaLF} corresponds to the width of the redshift bin.
The Ly$\alpha$ luminosity function at $\log{L_\mathrm{Ly\alpha}}=42-43~\mathrm{erg\ s^{-1}}$ decreases by $\sim3$ dex from $z\sim5$ to $z\sim10-14$, suggesting an increase in damping wing absorption.\par
We fit the Ly$\alpha$ luminosity function with the Schechter function form:
\begin{align}
    \Phi_\mathrm{Ly\alpha}&(L_\mathrm{Ly\alpha})\notag\\
    &=\ln{10}\phi^\star\qty(\frac{L_\mathrm{Ly\alpha}}{L^\star})^{\alpha+1}\exp\qty(-\frac{L_\mathrm{Ly\alpha}}{L^\star}).
\end{align}
We fix the $L^\star$ and $\alpha$ values at $z\sim5$, $6$, $7$, $8-9$, and $10-14$ to the ones at $z=4.9$, $5.7$, $7.0$, $7.3$, and $7.3$ derived by \citet{umeda24b}.
We obtain the posterior distribution of $\phi^\star$ with the MCMC method.
The best-fit parameters are presented in Table \ref{table:lyalf_fit}.

\section{Modeling Lyman-alpha EW Distributions}\label{sec;model}
In this section, we construct theoretical Ly$\alpha$ EW distribution models in order to estimate $\xHI$ by comparing them with the observed EW values.
We assume that \revise{the red side of} the Ly$\alpha$ emission line at $z\sim5$ is not absorbed by the IGM and that the intrinsic EW distribution at $z\gtrsim6$ is identical to the distribution at $z\sim5$.
The models of EW probability distribution are then calculated by multiplying the Ly$\alpha$ photon transmittance through the IGM (Section \ref{sec;sim}) to the intrinsic EW distribution (Section \ref{sec;int}).
\revise{Although we assume a constant intrinsic distribution, we discuss the potential impact of galaxy evolution in Section \ref{sec;history}.}

\subsection{Lyman-alpha EW Distribution at $z=5$}\label{sec;int}
We use the EW values of galaxies in the redshift range $4.5<z<5.5$ to estimate the intrinsic Ly$\alpha$ EW distribution.
For details about the derivation of the EW distribution, see Section \ref{sec;lyaLF}.
Previous studies have parameterized the intrinsic distribution using an exponential (e.g., \citealt{dijkstra11}), an exponential plus a delta function (e.g., \citealt{mason18}), or a log-normal distribution (e.g., \citealt{tang24a}).
In this work, we do not assume any specific functional form and instead numerically derive the distribution from the histogram of the observed EW values for each UV magnitude bin.
\revise{Although we construct the intrinsic Ly$\alpha$ EW distribution using galaxies at $4.5<z<5.5$, the choice of this redshift range does not significantly affect our results. Near the end of reionization, most galaxies reside in large ionized bubbles, making the Ly$\alpha$ EW distribution largely insensitive to $\xHI$ (see Figure \ref{ew_dist}). Even if a small fraction of neutral hydrogen remains (e.g., $\xHI \sim 0.05$ at $z=5$), its impact on the Ly$\alpha$ EW distribution is negligible compared to the uncertainty introduced by the limited sample size.}

\subsection{21cmFAST Simulation}\label{sec;sim}
\begin{figure}[t]
    \centering
    \includegraphics[width=\linewidth]{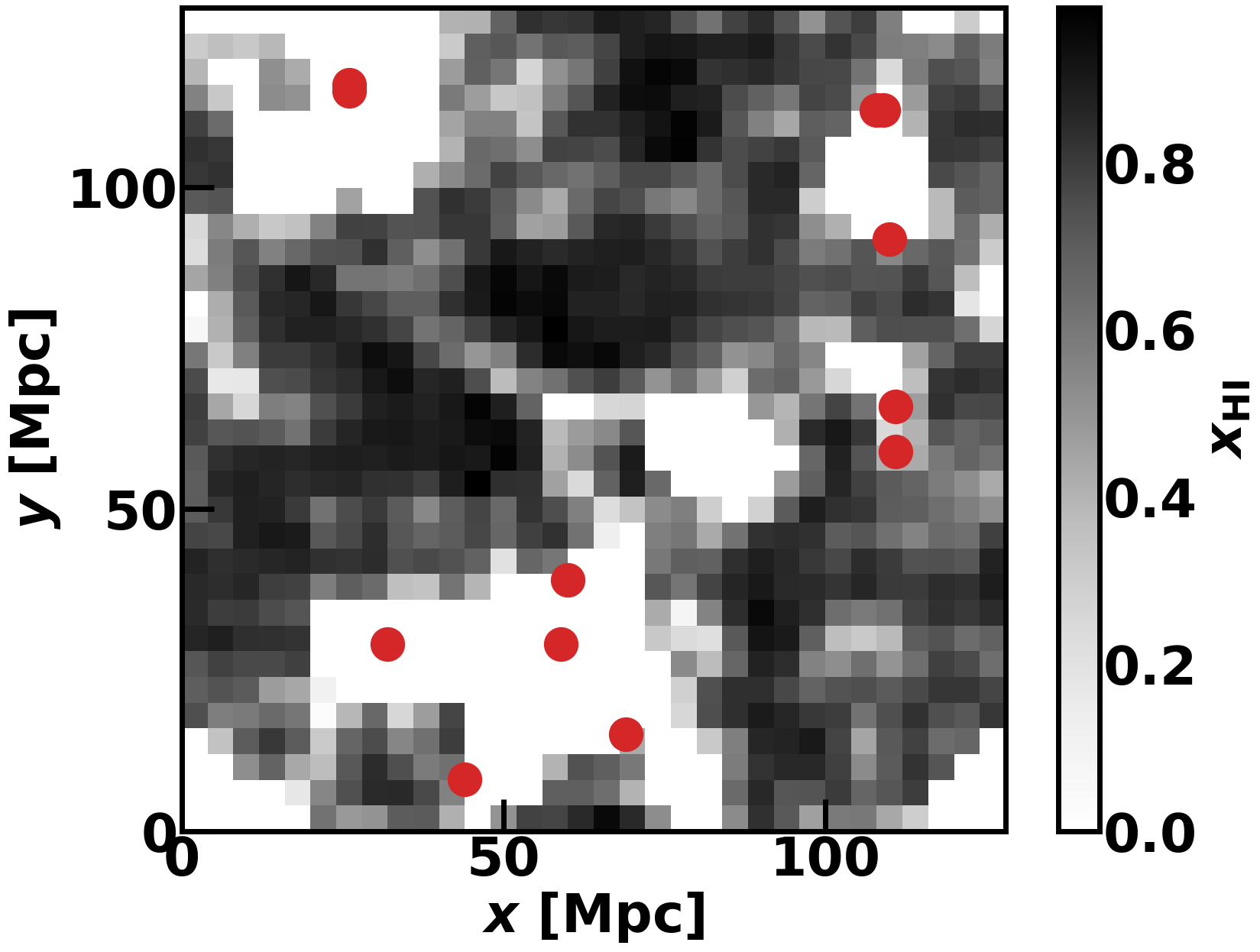}
    \caption{Spatial distribution of neutral regions (black) and halos (red) from our \texttt{21cmFAST} simulation box at $z=7$. While the full simulation uses a three-dimensional $1024^3~\mathrm{cMpc^3}$ box, this figure shows a two-dimensional slice of $128^2~\mathrm{cMpc^2}$.}
    \label{fastsim}
\end{figure}
\begin{figure}[t]
    \centering
    \includegraphics[width=\linewidth]{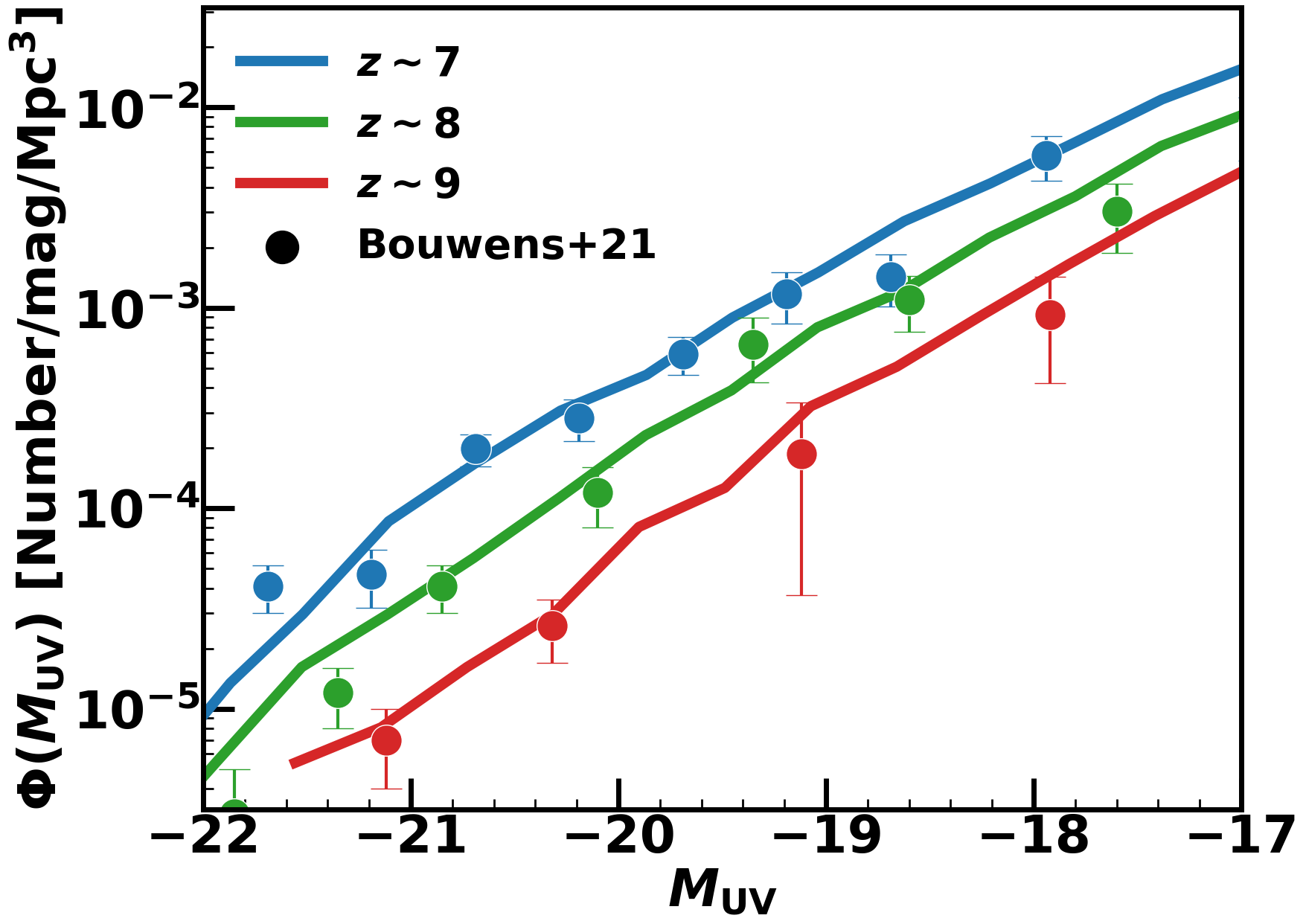}
    \caption{UV luminosity functions at \Add{$z=7$, $8$, and $9$. The blue, green, and red solid lines represent the UV luminosity functions at $z=7$, $8$, and $9$} derived from our \texttt{21cmFAST} simulation. \Add{The circles} show the observational UV luminosity functions reported by \citet{bouwens21}, which agree with our simulation results.}
    \label{UVLF}
\end{figure}
To model EW distributions at $z\gtrsim6$, we calculate the transmission fraction of Ly$\alpha$ photons \Add{following the methodology presented in \citet{mason18}}.
The optical depth of Ly$\alpha$ photons, $\tau$, depends on velocity $v$, redshift $z$, the volume-averaged neutral fraction $\xHI$, and the UV magnitude of the Ly$\alpha$-emitting galaxy $M_\mathrm{UV}$.
We assume that Ly$\alpha$ photons blueward of the circular velocity $v_c=(10GM_hH(z))^{1/3}$ are completely absorbed by the circumgalactic medium and the residual neutral gas within the ionized region in the IGM.
Here, $M_h$ is the halo mass.
We adopt the relation between $M_\mathrm{UV}$ and $M_h$ derived by \citet{mason15, mason18}:
\begin{align}\label{magmas}
    \log{M_h[M_\odot]}=\gamma(M_\mathrm{UV}+20.0+0.26z)+11.75,
\end{align}
where $\gamma=-0.3$ for $M_\mathrm{UV}>-20.0-0.26z$ and $\gamma=-0.7$ otherwise.\par
\revise{We assume that the optical depth is solely due to damping wing absorption in the IGM at $v>v_c$, where Ly$\alpha$ photons are expected to transmit efficiently after the EoR \citep{laursen11, byrohl20, gurunglopez20}.
Note that damped Ly$\alpha$ absorbers (DLAs) also contribute to absorption in reality.
Recent studies using large JWST archival datasets (e.g., \citet{heintz25, mason25}) find that a significant fraction of galaxies exhibit strong DLA absorption. However, they do not find a clear redshift evolution in this fraction. Given this, we do not explicitly model DLAs and instead attribute the redshift evolution of Ly$\alpha$ emission primarily to the IGM.}\par
\revise{To calculate the optical depth in the IGM,} we use the semi-numerical cosmological simulation code \texttt{21cmFAST v3} \citep{mesinger11, murray20} to model the IGM ionization and galaxy distributions.
We run a \texttt{21cmFAST} simulation from $z=15$ to $z=4$ with a box size of $1024^3~\mathrm{cMpc^3}$, a \Add{halo distribution map} resolution of 1~Mpc/cell, and an ionization box resolution of 4~Mpc/cell.
\Add{We calculate the IGM ionization and halo distribution maps from the same initial density map.}
Figure \ref{fastsim} illustrates a portion of our simulation box at $z=7$.\par
The reionization process in \texttt{21cmFAST} is primarily determined by two parameters: the ionizing efficiency $\zeta$ and the minimum virial temperature $T_\mathrm{vir}^\mathrm{min}$.
\revise{We explain the physical meaning of these parameters in the following.}
$\zeta$ represents the number of ionizing photons per baryon inside a halo. It is expressed as
\begin{align}\label{zeta}
    \zeta=N_\gamma f_\mathrm{esc}f_\star,
\end{align}
where $N_\gamma$ is the number of ionizing photons per stellar baryon, $f_\mathrm{esc}$ is the escape fraction of ionizing photons, and $f_\star$ is the fraction of stellar baryons relative to the total baryons in a halo.
\Add{Since the number of ionizing photons emitted from a halo is determined solely by $\zeta$, we do not assume specific values for $N_\gamma$ and $f_\star$.}
$T_\mathrm{vir}^\mathrm{min}$ defines the threshold temperature at which a halo can emit ionizing photons.
This temperature corresponds to the minimum halo mass capable of emitting ionizing photons $M_h^\mathrm{min}$, which is given by the relation \citep{barkana01}:
\begin{align}\label{temp_mass}
    T_\mathrm{vir}^\mathrm{min}&=1.98\times10^4\qty(\frac{\mu}{0.6})\notag\\
    &\times\qty(\frac{M_h^\mathrm{min}}{10^8h^{-1}M_\odot})^{2/3}\qty(\frac{\Omega_m\Delta_c}{\Omega_m^z18\pi^2})^{1/3}\qty(\frac{1+z}{10})~\mathrm{K},
\end{align}
where $\mu$ is the mean molecular weight, $\Omega_m^z=\frac{\Omega_m(1+z)^3}{\Omega_m(1+z)^3+1-\Omega_m}$, $\Delta_c=18\pi^2+82d-39d^2$, with $d=\Omega_m^z-1$.
In our simulation, we adopt fiducial parameters of $\zeta=20$ and $T_\mathrm{vir}^\mathrm{min}=5\times10^4~\mathrm{K}$.
We also set the mean free path of ionizing photons within ionized regions $R_\mathrm{mfp}$ to 15~cMpc to account for the effects of recombination.
\Add{In the \texttt{21cmFAST} simulations, $R_\mathrm{mfp}$ is treated as the maximum horizon of ionizing photons.}
\Add{For the determination of these fiducial parameters, see \citet{mesinger16} and \citet{greig17}}.\par
In our simulation box, we confirm that the UV luminosity function during the EoR and the CMB optical depth are reproduced.
We convert the halo mass function in our simulation box into the UV luminosity function using Equation (\ref{magmas}).
Figure \ref{UVLF} shows \Add{the UV luminosity functions at $z=7$, $8$, and $9$} from our simulation.
Our results are consistent with the measurements by \citet{bouwens21}.
\Add{The optical depth of the CMB, $\tau_e$, corresponding to the \texttt{21cmFAST} model reionization history, is calculated as:
\begin{align}\label{cmb}
    \tau_e=\int_0^{1090}&\dd{z}(1-\xHI(z))\frac{c(1+z)^2}{H(z)}\notag\\
    &\times n_0\sigma_\mathrm{T}\qty(1+\frac{\eta Y_p}{4X_p}),
\end{align}
where $\sigma_\mathrm{T}$ is the Thomson scattering cross section, $X_p=1-Y_p(=0.76)$ is the hydrogen abundance, and $\eta$ represents the relative ionization fraction of helium to hydrogen.
Following \citet{kuhlen12}, we assume helium is singly ionized ($\eta=1$) at $z>4$ and doubly ionized ($\eta=2$) at $z\leq4$.
The reionization history $\xHI(z)$ is obtained from the \texttt{21cmFAST} simulations.}
The obtained value, $\tau_e=0.054$, agrees with the Planck measurement ($\tau_e=0.0561\pm0.0071$; \citealt{planck20}).
Since our simulation reproduces key observations related to structure formation and reionization, it is well-suited for modeling the EW distribution during the EoR.\par
We calculate the optical depth of Ly$\alpha$ photons emitted from a halo \Add{with mass $M_h$ at redshift $z$} in the simulation box, \Add{as a function of} volume-averaged neutral fractions $\xHI$.
Since the reionization morphology has only a small effect when using galaxies spread in redshift ($\Delta z>0.1$ bin\Add{; e.g., \citealt{sobacchi15}}), we superimpose the ionization maps with different $\xHI$ values from a single simulation onto the halo distribution map. This approach follows the methodology of  \citet{mason18}.\par
Based on the formulation of \citet{miraldaescude98}, the optical depth is calculated using the following equation:
\begin{align}
    \tau&(v,z,\xHI,M_\mathrm{UV}(M_h))\notag\\
    &=\int_5^z\frac{\dd{z'}}{1+z'}\frac{c}{H(z')}n_0(1+z')^3X_{\rm H\,\textsc{i}}(z'\Add{, \xHI, M_\mathrm{UV}})\notag\\
    &\ \ \ \times\sigma\qty(\frac{\omega}{\omega_a}=\frac{1+z'}{(1+z)(1+v/c)}).
\end{align}
Here, we assume that reionization has ended at $z=5$.
The halo mass $M_h$ is converted into $M_\mathrm{UV}$ using Equation (\ref{magmas}).
$X_{\rm H\,\textsc{i}}(z\Add{, \xHI, M_\mathrm{UV}})$ is the line-of-sight neutral fraction distribution of the halo in the simulation box, which differs from the volume-averaged neutral fraction \Add{$\xHI$} at redshift $z$.
\Add{Although \citet{miraldaescude98} assumes a constant $\xHI$ value, we consider the effect of inhomogeneous neutral hydrogen distributions.}
\Add{The value} $n_0$ represents the comoving hydrogen number density.
$\sigma(\omega)$ is the Ly$\alpha$ damping wing cross section at the velocity $v=\omega/(2\pi)$, given by
\begin{align}
    \sigma(\omega)=\frac{3\lambda_a^2\Lambda^2}{8\pi}\frac{(\omega/\omega_a)^4}{(\omega-\omega_a)^2+(\Lambda^2/4)(\omega/\omega)^6},
\end{align}
where $\omega_a=2\pi c/\lambda_a$ \Add{and} $\Lambda=6.25\times10^8~\mathrm{s^{-1}}$.\par
Using the optical depth, we calculate the transmission fraction \Add{$\mathcal{T}$}.
Following \citet{mason18}, we \Add{model the Ly$\alpha$ lineshape after transmission through the ISM as a Gaussian}, \Add{with the full width at half maximum (FWHM) equal to the peak velocity offset $\Delta v$}.
\Add{We use the value of $\Delta v$ which} follows a halo mass-dependent \Add{log-normal} distribution described in \citet{mason18}, \Add{whose peak is expressed as:
\begin{align}\label{vel_cen}
    \log\Delta v_\mathrm{peak}(M_h)=0.32\log\qty(\frac{M_h}{1.55\times10^{12}M_\odot})+2.48.
\end{align}
}
The transmission fraction is then calculated using the equation:
\begin{align}
    \mathcal{T}&(z,\xHI,M_\mathrm{UV}, \Delta v)\notag\\
    &=\frac{\int_{v_c}^\infty\dd{v}\exp\qty(-\frac{(v-\Delta v)^2}{2\sigma_a^2})\exp(-\tau(v,z,\xHI,M_\mathrm{UV}))}{\int_{v_c}^\infty\dd{v}\exp\qty(-\frac{(v-\Delta v)^2}{2\sigma_a^2})},
\end{align}
where $2\sqrt{2\ln{2}}\sigma_a=\Delta v$.
Because $\mathcal{T}$ varies significantly across sightlines, \Add{we} calculate $\mathcal{T}$ for all the halos in the simulation box to derive the transmission fraction as a probability distribution $p(\mathcal{T}\mid z,\xHI,M_\mathrm{UV}, \Delta v)$.

\subsection{Lyman-alpha EW Distribution at $z\gtrsim6$}\label{sec;ewobs}
\begin{figure}[t]
    \centering
    \includegraphics[width=\linewidth]{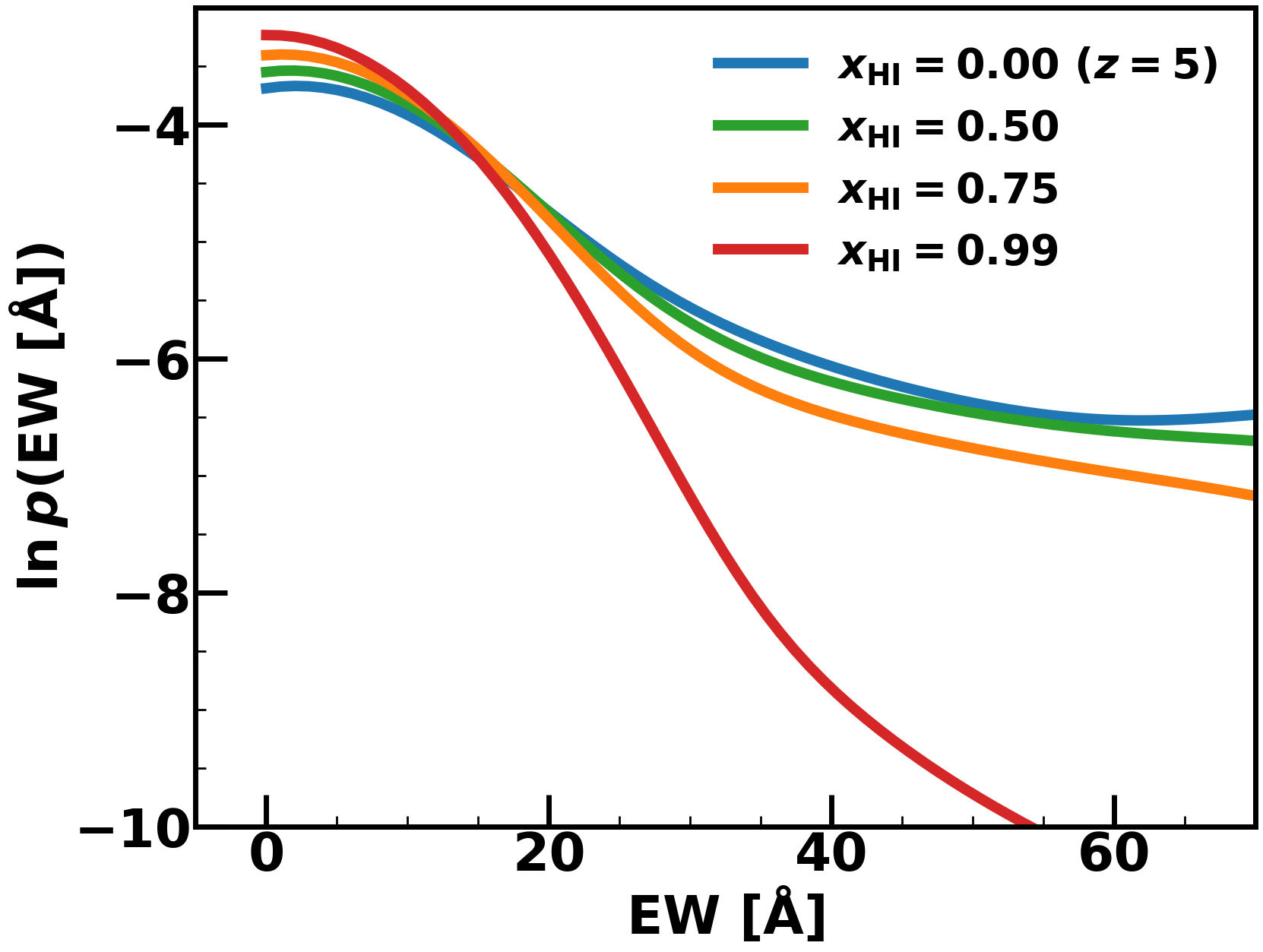}
    \caption{\Add{Models of} Ly$\alpha$ EW distributions for galaxies with $M_\mathrm{UV}\sim-18.5$ at $z=7$. We show the distributions for four values of $\xHI$. The distribution for $\xHI=0.00$ represents the intrinsic distribution at $z=5$ without IGM absorption. As $\xHI$ increases, the probability of observing large EW values decreases due to stronger absorption by the IGM.}
    \label{ew_dist}
\end{figure}
By combining the \Add{observed} intrinsic EW distribution with the \Add{simulated} transmission distribution, we construct \Add{models of} EW distributions at $z\gtrsim6$ as a function of $\xHI$ and $M_\mathrm{UV}$ through the following steps.
First, an intrinsic EW value $\mathrm{EW_{int}}$ is randomly selected from the distribution derived in Section \ref{sec;int}.
Next, the velocity offset is randomly determined from the distribution $p(\Delta v\mid M_\mathrm{UV})$.
Using this velocity offset, the transmission fraction $\mathcal{T}$ is randomly chosen from the distribution $p(\mathcal{T}\mid z,\xHI,M_\mathrm{UV}, \Delta v)$ derived in Section \ref{sec;sim}.
The observed EW value is then calculated as $\mathrm{EW_{obs}}=\mathrm{EW_{int}}\mathcal{T}$. 
By repeating this procedure multiple times, we obtain the observed EW distribution $p(\mathrm{EW}\mid z,\xHI,M_\mathrm{UV})$.
In Figure \ref{ew_dist}, we present examples of the distributions for galaxies with $M_\mathrm{UV}\sim-18.5$ at $z=7$. As $\xHI$ increases, Ly$\alpha$ photons experience greater absorption, reducing the likelihood of observing large EW values.

\section{Observational Constraints on the Cosmic Reionization History}\label{sec;crh}
\subsection{Neutral Fraction Estimate}\label{sec;xhi_estimate}
\begin{figure}[t]
    \centering
    \includegraphics[width=0.8\linewidth]{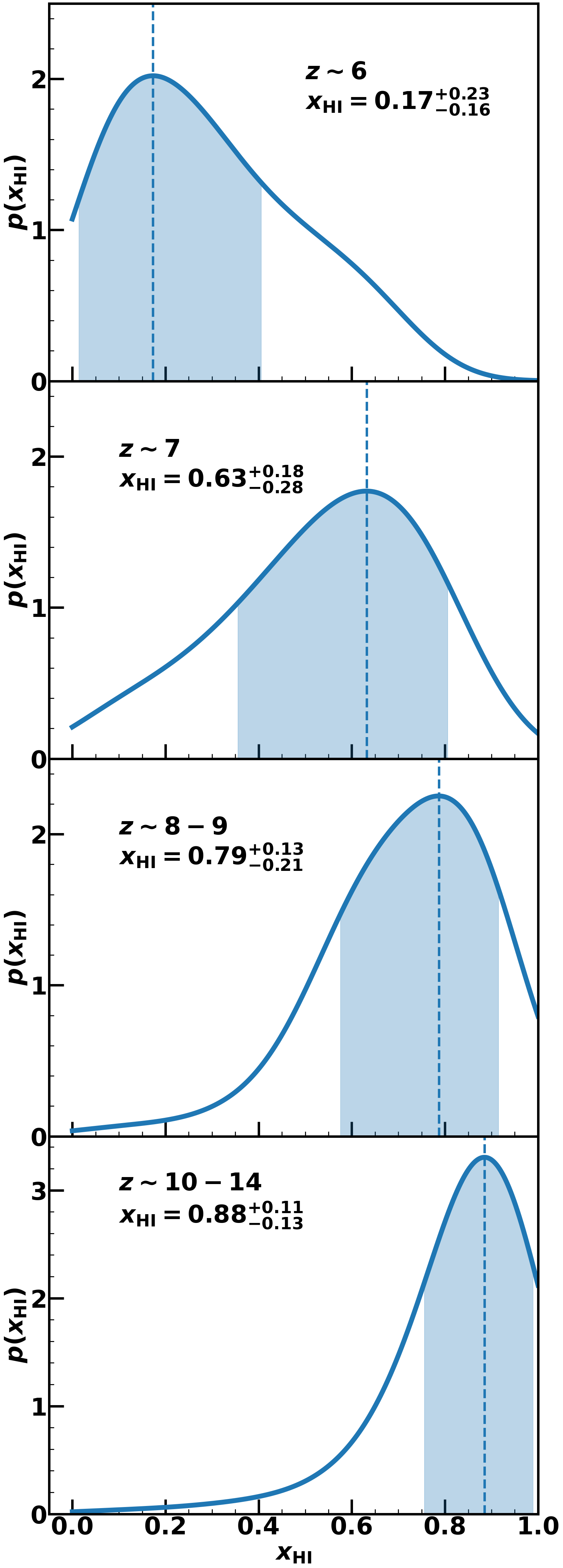}
    \caption{Posterior distributions of $\xHI$ at $z\sim6$, $7$, $8-9$, and $10-14$ (from top to bottom). The vertical dashed lines represent the mode of each distribution, and the shaded regions indicate the $68\%$ HPDI. The estimated values are $\xHI=0.17_{-0.16}^{+0.23}$, $0.63_{-0.28}^{+0.18}$, $0.79_{-0.21}^{+0.13}$, and $0.88_{-0.13}^{+0.11}$, respectively.}
    \label{xhi_pos}
\end{figure}
By comparing the observed EW values to the models of EW distributions, we estimate the neutral fraction $\xHI$.
We divide $z>5.5$ galaxies in our sample into four redshift bins: $5.5<z\leq6.5$, $6.5<z\leq7.5$, $7.5<z\leq9.5$, and $9.5<z\Add{\leq14.2}$.
The number of galaxies in each bin is \Add{159, 128, 42, and 21}, respectively.
At each redshift bin, the posterior distribution of $\xHI$ is derived using Bayes' theorem:
\begin{align}
    p_z&(\xHI\mid\{\mathrm{EW},M_\mathrm{UV}\})\notag\\
    &\propto\prod_ip(\mathrm{EW}_i\mid z,\xHI,M_{\mathrm{UV},i}))p(\xHI),
\end{align}
where $\mathrm{EW}_i$ and $M_{\mathrm{UV},i}$ are the EW value and UV magnitude of the $i$th galaxy\Add{, respectively}.
\Add{We assume a flat prior distribution for $\xHI$.}
Given that the EW value of each galaxy contains an error from spectral fitting, the equation is rewritten as:
\begin{align}
    p_z&(\xHI\mid\{\mathrm{EW},M_\mathrm{UV}\})\notag\\
    &\propto\prod_i\int\dd{\mathrm{EW}}p(\mathrm{EW}\mid z,\xHI,M_{\mathrm{UV},i})p_i(\mathrm{EW}),
\end{align}
where $p_i(\mathrm{EW})$ is the posterior EW distribution of the $i$th galaxy obtained from spectral MCMC fitting, and $p(\mathrm{EW}\mid z,\xHI,M_{\mathrm{UV},i})$ is the EW distribution model derived in Section \ref{sec;ewobs}.
The posterior distributions of $\xHI$ for each redshift bin are shown in Figure \ref{xhi_pos}.
We determine the $\xHI$ values and their $1\sigma$ errors by the mode and $68\%$ highest posterior density interval (HPDI) of the posterior distributions. The estimated values are $\xHI=0.17_{-0.16}^{+0.23}$, $0.63_{-0.28}^{+0.18}$, $0.79_{-0.21}^{+0.13}$, and $0.88_{-0.13}^{+0.11}$ at $z\sim6$, $7$, $8-9$, and $10-14$, respectively (Table \ref{table:xhi}).
\begin{deluxetable}{cccc}
    \tablecolumns{2}
    \tablewidth{\linewidth}
    \tabletypesize{\scriptsize}
    \tablecaption{$\xHI$ Estimate
    \label{table:xhi}}
    \tablehead{
    \colhead{$\expval{z}$} & \colhead{$z$ range} & \colhead{$N_\mathrm{gal}$} & \colhead{$\xHI$}\\
    (1) & (2) & (3) & (4)
    }
    \startdata
    $5.90$ & $5.50\leq z\leq6.39$ & $159$ & $0.17_{-0.16}^{+0.23}$ \\
    $6.96$ & $6.54\leq z\leq7.49$ & $129$ & $0.63_{-0.28}^{+0.18}$ \\
    $8.41$ & $7.51\leq z\leq9.43$ & $42$ & $0.79_{-0.21}^{+0.13}$ \\
    $11.00$ & $9.62\leq z\leq14.18$ & $21$ & $0.88_{-0.13}^{+0.11}$
    \enddata
    \tablecomments{(1): Mean redshift. (2) Redshift range. (3) Number of galaxies in a subsample. (4): Estimated $\xHI$ value at each redshift. The mode and 68\% HPDI are presented.}
\end{deluxetable}

\subsection{Cosmic Reionization History}\label{sec;history}
\begin{figure*}[t]
    \centering
    \includegraphics[width=\linewidth]{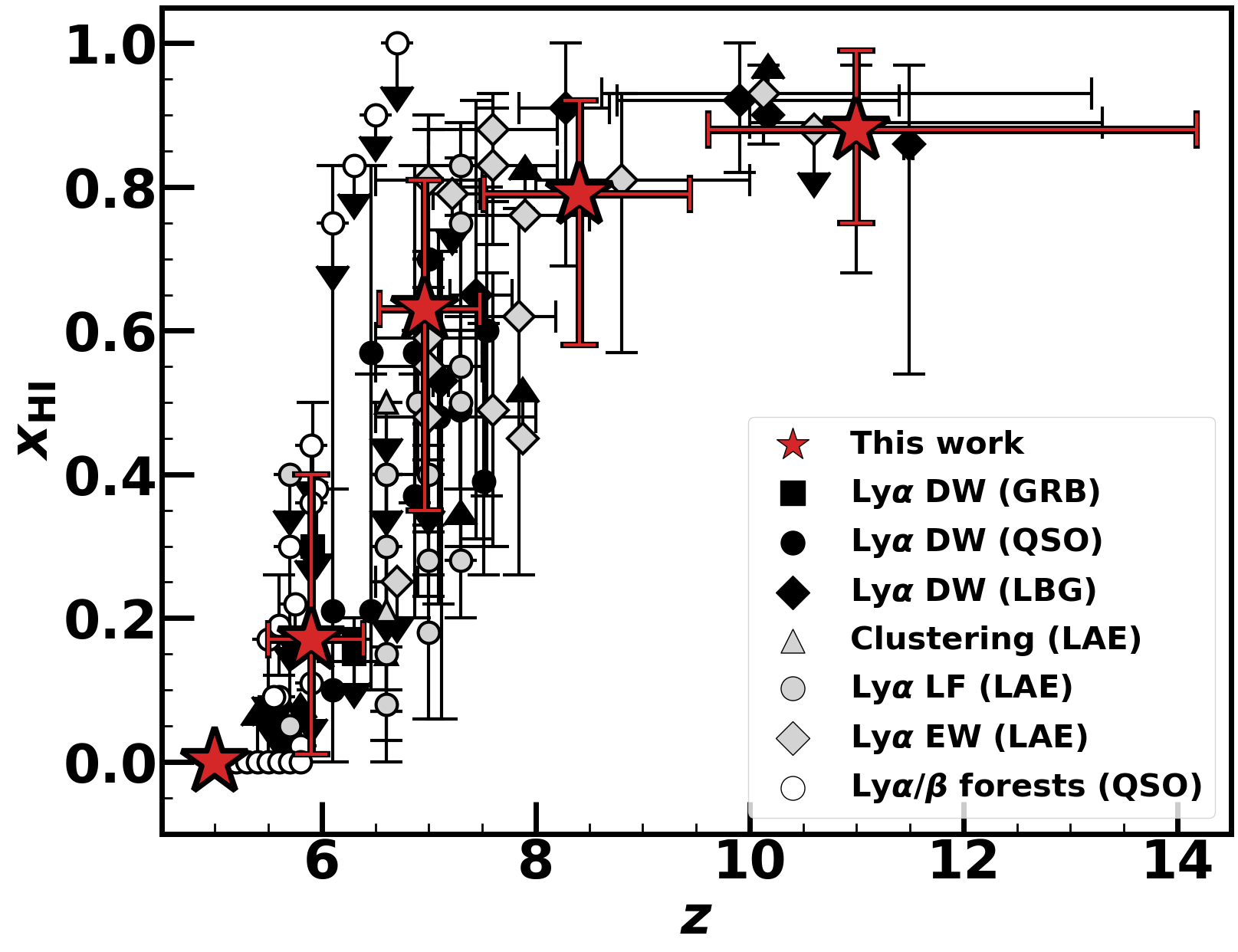}
    \caption{Redshift evolution of $\xHI$. The red stars at $z\gtrsim6$ represent the $\xHI$ values obtained in this work using Ly$\alpha$ EW distributions. At $z=5$, $\xHI$ is assumed to be zero in this study. We also present $\xHI$ estimates from the literature using various methods: Ly$\alpha$ damping wing absorption of GRBs (black squares; \citealt{totani06, totani14, fausey24}), QSOs (black circles; \citealt{davies18, wang20, yang20, greig22, durovcikova24}), and LBGs (black diamonds; \citealt{curtislake23, hsiao24, umeda24a}); LAE clustering (gray triangles; \citealt{sobacchi15, ouchi18, umeda24b}); Ly$\alpha$ luminosity function (gray circles; \citealt{ouchi10, konno14, konno18, zheng17, inoue18, morales21, goto21, ning22, umeda24b}); Ly$\alpha$ EW distribution (gray diamonds; \citealt{mason18, mason19, hoag19, jung20, whitler20, bolan22, bruton23, morishita23, nakane24, tang24b, jones24}); and Ly$\alpha$ forests and/or Ly$\alpha+\beta$ dark fraction/gaps (white circles; \citealt{mcgreer15, bosman22, zhu22, zhu24, jin23, spina24}).}
    \label{crh}
\end{figure*}
\begin{figure}
    \centering
    \includegraphics[width=0.9\linewidth]{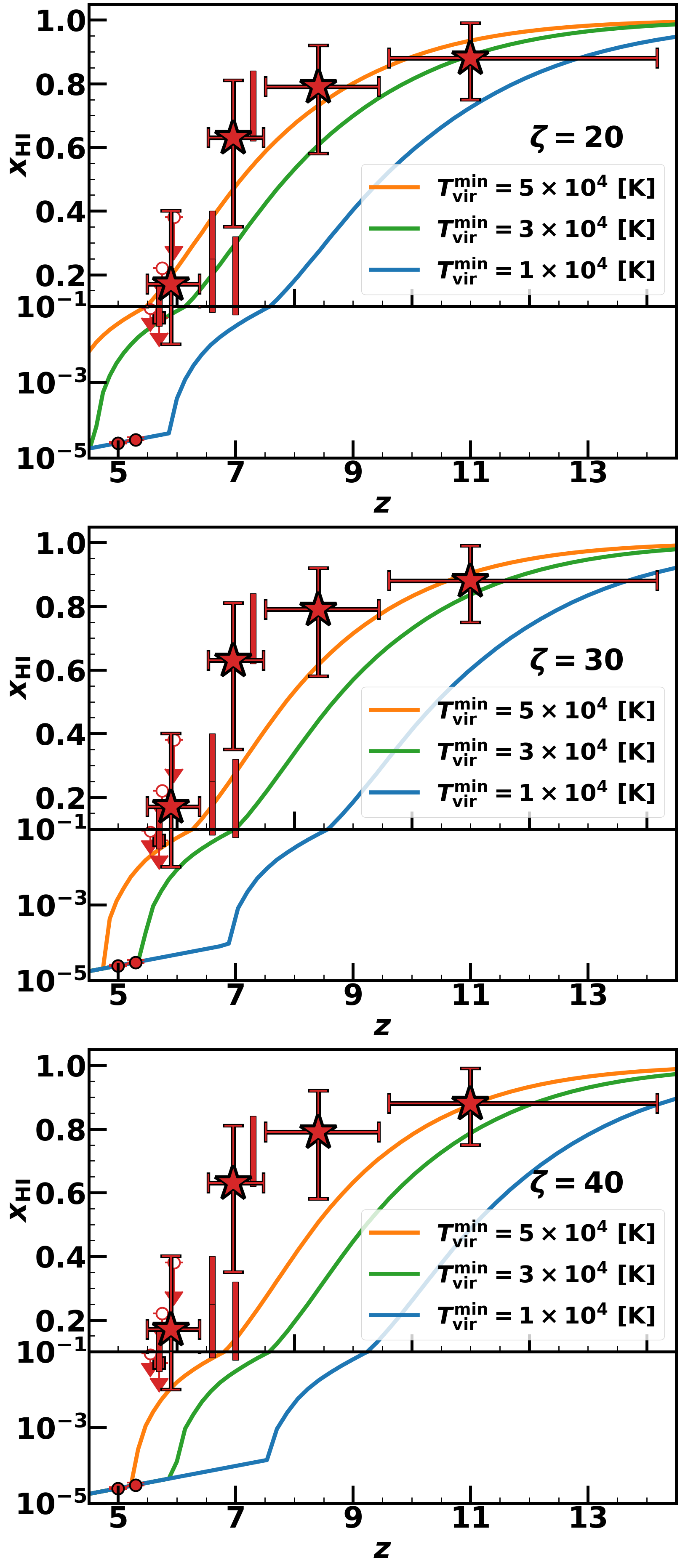}
    \caption{Models of cosmic reionization history. The curves show the results of nine \texttt{21cmFAST} simulations with different combinations of $\zeta$ and $T_\mathrm{vir}^\mathrm{min}$. The red stars indicate the $\xHI$ values derived in this work. The red rectangles, open circles, and red circles show observational constraints from Ly$\alpha$ luminosity function/angular correlation function \citep{umeda24b}, Ly$\alpha+\beta$ forest dark gaps \citep{zhu22}, and Ly$\alpha$ forest optical depth \citep{bosman22}, respectively. The observational results suggest a sharper reionization history than any of the nine models.}
    \label{crh_model}
\end{figure}
Figure \ref{crh} shows our $\xHI$ constraints at $z=6-14$.
Our high $\xHI$ values suggest a late reionization.
For comparison, we include $\xHI$ estimates from the literature.\par
Our $\xHI$ value at $z\sim6$ is consistent with estimates from Ly$\alpha$ damping wing absorption of GRBs and QSOs, LAE clustering, Ly$\alpha$ luminosity function, and Ly$\alpha/\beta$ forests (e.g., \citealt{totani14, durovcikova24, umeda24b, spina24}).
At higher redshift ($z\sim8-9$ and $10-14$), our $\xHI$ values also agree with estimates from Ly$\alpha$ EW distributions and Ly$\alpha$ damping wing absorption of LBGs (e.g., \citealt{bruton23, curtislake23, hsiao24, nakane24, tang24b, umeda24a}). \Add{Note that} many of the galaxies used in these studies are included in our sample.
At $z\sim7$, numerous $\xHI$ estimates exist, some of which are inconsistent with each other.
Our moderately high neutral fraction at $z\sim7$ ($\xHI=0.63_{-0.28}^{+0.18}$) is consistent with many other measurements (e.g., \citealt{durovcikova24, nakane24, tang24b}), with some studies showing even higher best-fit values (\citealt{wang20}: $\xHI=0.70_{-0.23}^{+0.20}$ and \citealt{jones24}: $\xHI=0.81_{-0.10}^{+0.07}$).
\revise{It is also consistent with reionization history obtained by \citet{choustikov25} based on the \texttt{SPHINX} simulation and JWST NIRCam photometric measurements.}
However, our result is inconsistent with some estimates from Ly$\alpha$ luminosity functions (\citealt{morales21}: $\xHI=0.28\pm0.05$ and \citealt{umeda24b}: $\xHI=0.18_{-0.12}^{+0.14}$ at $z=7.0$).
Interestingly, \citet{morales21} and \citealt{umeda24b} both report high $\xHI$ at $z=7.3$ (\citealt{morales21}: $\xHI=0.83_{-0.07}^{+0.06}$ and \citealt{umeda24b}: $\xHI=0.75_{-0.13}^{+0.09}$), suggesting a very sharp reionization history from $z=7.3$ to $z=7.0$.
These results share the late reionization feature found in our study, although the timing of the sharp progress differs slightly.
This inconsistency might be attributed to differences in redshift binning.
Note that our $z\sim7$ bin has a width of $\Delta z\sim1$, so galaxies at $z=7.0$ and $z=7.3$ are both included in the $z\sim7$ bin.\par
We compare our observational results to models of reionization history.
We use \texttt{21cmFAST} to generate reionization scenarios.
We perform a total of nine simulations by varying two input parameters, $\zeta$ and $T_\mathrm{vir}^\mathrm{min}$, each in three different ways.
The values of $\zeta$ are set to 20, 30, and 40, while $T_\mathrm{vir}^\mathrm{min}$ is set to $1\times10^4$, $3\times10^4$, and $5\times10^4$ K.
Figure \ref{crh_model} represents these nine reionization scenarios.
In all scenarios, the high observed $\xHI$ values at $z\geq7$ cannot be explained, or reionization does not complete by $z=5.3$.
When $\zeta=40$, the ionizing efficiency is too high, causing reionization to finish too early, which is inconsistent with the observational results.
For $\zeta=30$, increasing $T_\mathrm{vir}^\mathrm{min}$ brings the model closer to the observations at $z\geq7$.
However, it becomes inconsistent with the results of \citet{bosman22}, which suggest that reionization is almost complete at $z=5.3$ (i.e., $\xHI\sim10^{-5}$).
Similarly, for $\zeta=20$, it becomes easier to explain late reionization at $z\geq7$, but the value of $\xHI$ remains too high at $z=5.3$.
Therefore, a very sharp reionization scenario (i.e., reionization progresses rapidly around $z\sim6-7$) is required to satisfy the high $\xHI$ values observed at $z\geq7$ while ensuring that reionization completes by $z=5.3$.
Note that the resolution of \texttt{21cmFAST} does not provide sufficient accuracy to precisely determine the end of reionization (i.e., $\xHI\sim10^{-5}$).
Therefore, once the model reaches the extrapolation of the results by \citet{bosman22} at $z=5.0-5.3$, the reionization history follows the extrapolated values in Figure \ref{crh_model}.
Although this section compares the observational results only with several models, we perform fitting to the observational data by treating $\zeta$ and $T_\mathrm{vir}^\mathrm{min}$ as free parameters in Section \ref{sec;massivehalos}.
\par
\revise{An important point to consider is that our $\xHI$ estimate at $z\sim7$ has a relatively large uncertainty. The uncertainty in $\xHI$ increases when the variation in IGM absorption is large or when IGM absorption is less sensitive to $\xHI$. At the early stages of reionization, ionized bubbles around galaxies are small, and reionization progresses relatively homogeneously. This creates a strong correlation between IGM absorption and $\xHI$, leading to a more constrained estimate. However, during the middle stages of reionization, the ionization distribution becomes highly inhomogeneous. Some galaxies reside in large ionized bubbles with negligible IGM absorption, while others are in smaller bubbles and experience strong Ly$\alpha$ attenuation. This inhomogeneity contributes to increased uncertainty in $\xHI$. In the final stages of reionization, most observable galaxies are located within large ionized regions, making IGM absorption largely insensitive to $\xHI$, which further limits our ability to precisely constrain its value.}\par
\Add{Note that we estimate $\xHI$ assuming that the intrinsic Ly$\alpha$ EW distribution does not evolve with redshift. 
Some studies suggest that the intrinsic Ly$\alpha$ EW distribution shifts toward larger EWs at higher redshift (e.g., \citealt{naidu22}) due to the increasing fraction of young, metal-poor galaxies.
If this is the case, an even later reionization history would be required to explain the decreasing trend of observed Ly$\alpha$ EWs.}

\subsection{Onset of Cosmic Reionization}
\Add{Recently, a faint ($M_\mathrm{UV}=-18.7$) LAE with a Ly$\alpha$ EW of 43~\AA\ at $z=13.0$, JADES-GS-z13-1-LA, was identified by \citet{witstok24}.
We include this object in our sample.
We test whether the existence of such a strong LAE is consistent with our high $\xHI$ value, $\xHI\sim0.88$, at $z\sim10-14$.
To address this, we construct a Ly$\alpha$ EW distribution model at $z=13.0$ with $\xHI=0.88$ and $M_\mathrm{UV}=-18.7$, following the methodology described in Section \ref{sec;ewobs}.
In this model, the probability of observing a galaxy with $\mathrm{EW}>43$~\AA\ is 5\%.
Given that our sample includes five galaxies with $M_\mathrm{UV}\geq-18.7$ at $z=10-14$, the existence of JADES-GS-z13-1-LA is naturally explained within our framework.
Although our simulation does not fully resolve the complex morphology of ionized bubbles at the onset of reionization, \citet{qin24a} have investigated the detectability of Ly$\alpha$ emission at $z=13$ using more sophisticated simulations. Their results suggest that the probability of observing Ly$\alpha$ emission with $\mathrm{EW}>40$~\AA\ is $\sim10\%$.
Thus, our high $\xHI$ value at $z=10-14$, based on Ly$\alpha$ emission from galaxies including JADES-GS-z13-1-LA, is self-consistent.}

\section{Discussion}\label{sec;discussion}
In this section, we interpret the late and sharp reionization history derived in Section \ref{sec;crh}.
One possibility is the significant impact of recombination.
Recent observations have \Add{reported} that the mean free path of ionizing photons at $z\sim6$ is relatively short (e.g., \citealt{becker21, gaikwad23, zhu23, davies24a, satyavolu24}).
\Add{These findings} may imply the existence of more small-scale structures in the IGM than accounted for in traditional reionization simulations, resulting in a high clumping factor (\Add{$C\sim12$}; \citealt{davies24b}).
Under these conditions, the formation of large ionized bubbles may be hindered by recombination until the galaxy number density becomes sufficiently high at later epochs, potentially explaining the observed late reionization history.
Another possible interpretation is that massive halos are the primary sources of reionization.
Within the framework of hierarchical structure formation, late reionization can be explained by the delayed formation of massive halos, which emerge as the dominant contributors at later epochs (e.g., \citealt{naidu20}).
In Section \ref{sec;recombination}, we examine the role of strong recombination.
In Section \ref{sec;massivehalos}, we explore the scenario in which massive halos drive reionization.
Section \ref{sec;sum_dis} summarizes our discussion.

\subsection{Late Reionization by Strong Recombination}\label{sec;recombination}
\begin{figure}[t]
    \centering
    \includegraphics[width=\linewidth]{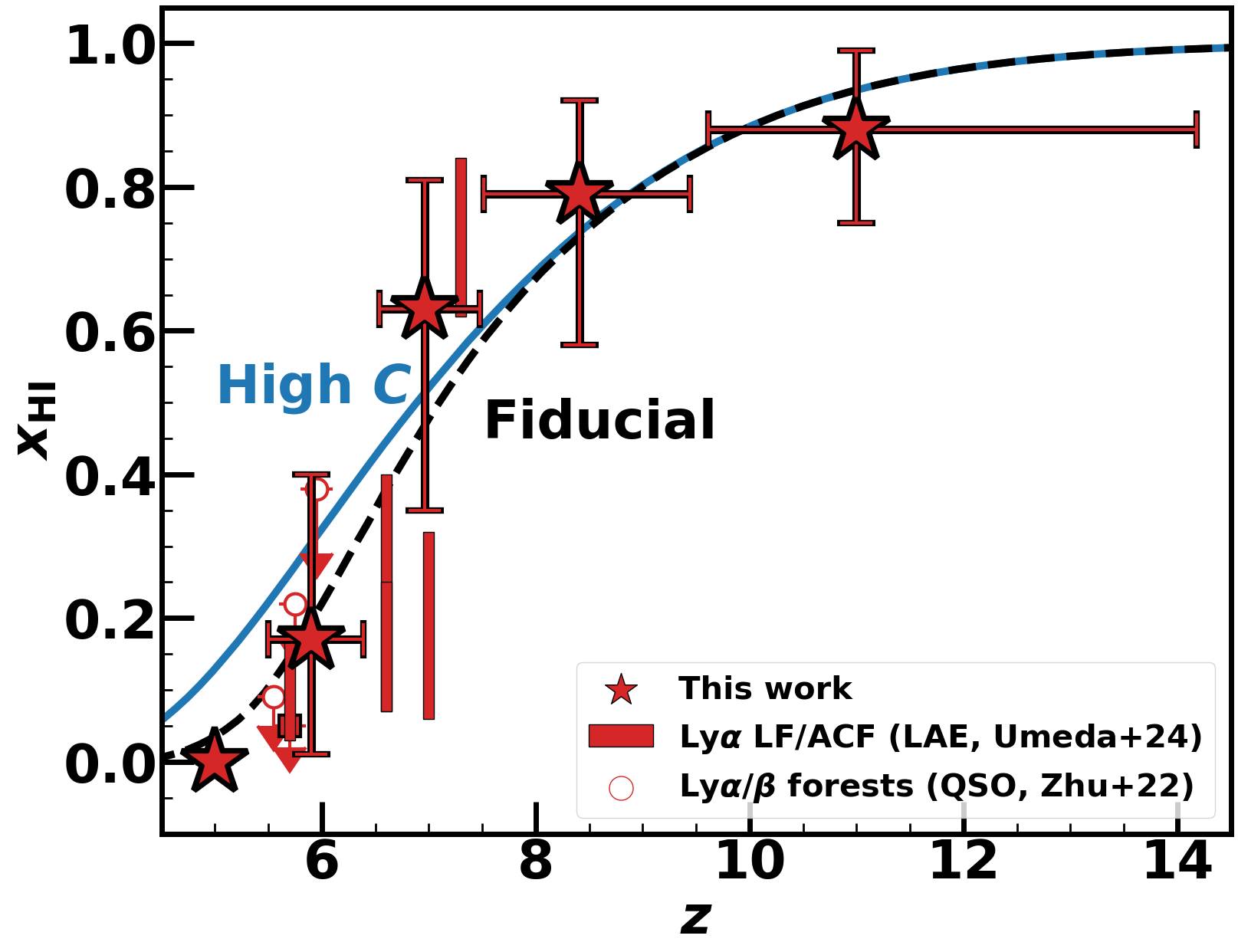}
    \caption{Reionization scenario assuming a high recombination rate. \Add{The blue solid line shows the reionization history with a short mean free path ($R_\mathrm{mfp}=6$~cMpc), which corresponds to a high clumping factor $C$.} The black dashed line represents the history with fiducial parameters ($R_\mathrm{mfp}=15$~cMpc), which corresponds to a low clumping factor. The scenario with a short mean free path shows a more gradual $\xHI$ evolution than the fiducial one.}
    \label{mfp}
\end{figure}
In this section, we focus on the impact of strong recombination.
Observations of quasar spectra by \citet{zhu23} indicate that the mean free path of ionizing photons at $z=5.93$ is $\sim0.8$~pMpc\Add{, corresponding to a clumping factor of $C\sim12$ \citep{davies24b}}.
To match this result, we set the mean free path in the \texttt{21cmFAST} simulation to 6~cMpc, treating it as a redshift-independent parameter.
We adopt the default values of $\zeta=20$ and $T_\mathrm{vir}^\mathrm{min}=5\times10^4$ K, which \Add{corresponds to a limiting magnitude} $M_\mathrm{UV}=-10$ \citep{das17, greig17}\Add{, in order to account for the contribution from faint galaxies ($M_\mathrm{UV}\gtrsim-17$)}.\par
The blue solid and black dashed lines in Figure \ref{mfp} represent the scenario with $R_\mathrm{mfp}=6$~cMpc and $R_\mathrm{mfp}=15$~cMpc, respectively.
The scenario with a short mean free path (i.e., a high clumping factor) shows a more gradual decline in $\xHI$ at $z\sim6-7$ compared to the fiducial case ($R_\mathrm{mfp}=15$~cMpc), highlighting the difficulty of achieving a sharp reionization history when assuming a short mean free path.
\citet{qin24b} simulate the reionization history using the constraints from the CMB, UV luminosity functions, and forest data (without high-z LAE observations), and found a similar gradual evolution. \Add{Although a high clumping factor (or a short mean free path) naturally explains the residual neutral islands at $z\lesssim6$, the sharp reionization history is difficult to explain with this effect.}

\subsection{Reionization Driven by Massive Halos}\label{sec;massivehalos}
\begin{figure}[t]
    \centering
    \includegraphics[width=\linewidth]{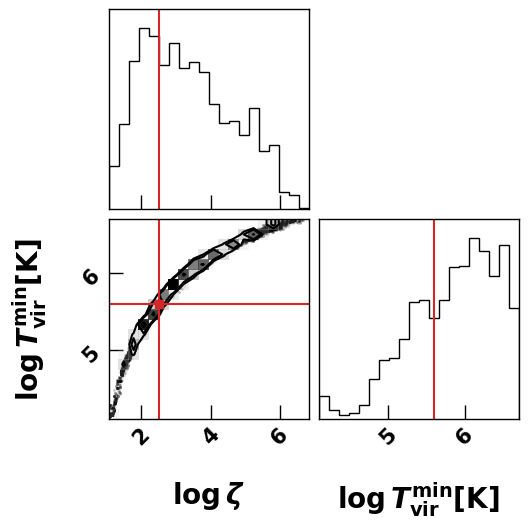}
    \caption{Posterior distributions of EoR parameters. The plot shows the ionizing efficiency $\zeta$ and the minimum virial temperature required for a halo to emit ionizing photons $T_\mathrm{vir}^\mathrm{min}$. The red lines indicate the best-fit values: $\log{\zeta}=2.5$ and  $\log{T_\mathrm{vir}^\mathrm{min}/\mathrm{K}}=5.6$. Such a high $T_\mathrm{vir}^\mathrm{min}$ corresponds to the minimum halo mass $M_h^\mathrm{min}\sim10^{10.5}M_\odot$ at $z\sim6.5$.}
    \label{eor_params}
\end{figure}
\begin{figure}[t]
    \centering
    \includegraphics[width=\linewidth]{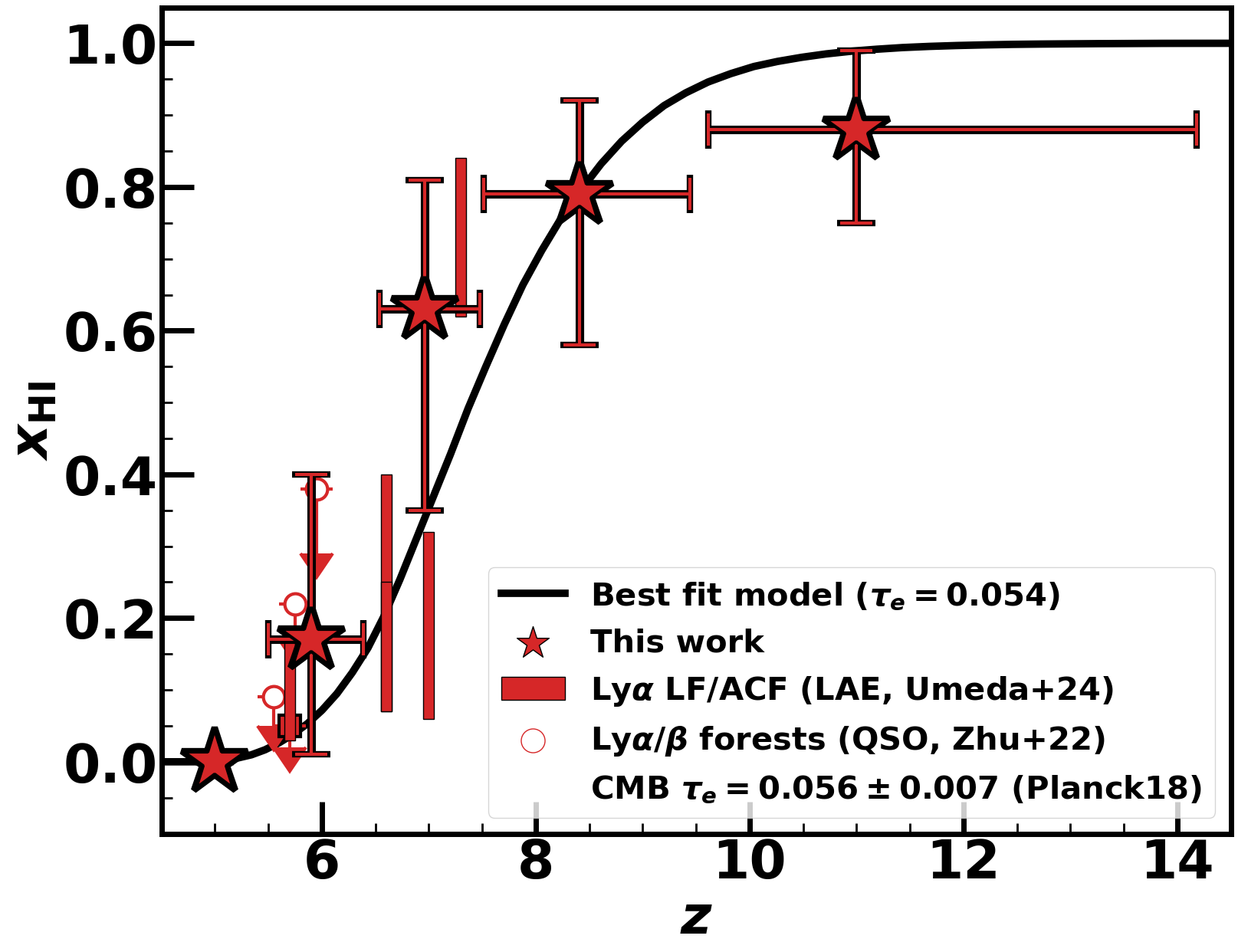}
    \caption{Cosmic reionization history: observational constraints and best-fit model. The black line represents the best-fit reionization scenario obtained from the parameter fitting procedure. The red stars indicate the $\xHI$ values derived in this work. \Add{The red rectangles and open circles show observational constraints from Ly$\alpha$ luminosity function/angular correlation function \citep{umeda24b} and Ly$\alpha+\beta$ forest dark gaps \citep{zhu22}, respectively, which are used for the fitting procedure}. \revise{We do not plot the CMB optical depth in this figure, as it does not correspond to a specific redshift $\xHI$ value. However, it is also included in the fitting procedure.} The best-fit scenario reproduces the observed $\xHI$ values and the CMB optical depth, supporting a late and sharp reionization.}
    \label{fit_crh}
\end{figure}
To explain the $\xHI$ measurements indicating a late and sharp reionization history, we fit observational constraints on $\xHI$, including our results, to reionization scenarios generated with \texttt{21cmFAST}.
We treat the ionizing efficiency $\zeta$ and the minimum virial temperature $T_\mathrm{vir}^\mathrm{min}$ as free parameters (see Section \ref{sec;sim}).
Each \texttt{21cmFAST} simulation run with these two EoR parameters provides a redshift evolution of $\xHI$, which we compare with observational results to estimate the best-fit parameters.
A high $T_\mathrm{vir}^\mathrm{min}$ corresponds to reionization dominated by massive halos, as it sets the minimum halo mass capable of emitting ionizing photons.
Since \texttt{21cmFAST} can incorporate inhomogeneous reionization processes, we use observational results reflecting such processes for our fitting.
Specifically, the following constraints are applied:
\begin{itemize}
    \setlength{\itemsep}{0pt}
    \item Ly$\alpha$ EW distributions observed by JWST (this work)
    \item Ly$\alpha$ luminosity functions and angular correlation functions observed by Subaru \citep{umeda24b}
    \item Ly$\alpha$ and Ly$\beta$ forest dark gaps of QSOs observed by Keck and VLT \citep{zhu22}
    \item Optical depth of CMB observed by Planck ($\tau_e=0.0561\pm0.0071$; \citealt{planck20}).
\end{itemize}
\Add{We calculate the optical depth of the CMB using Equation (\ref{cmb}). We adopt flat priors of $-\infty<\log{\zeta}<\infty$ and $4.0<\log{T_\mathrm{vir}^\mathrm{min}/\mathrm{K}}<6.7$.
The lower limit $\log{T_\mathrm{vir}^\mathrm{min}}=4.0$ corresponds to the threshold for efficient atomic cooling \citep{barkana01}, while the upper limit $\log{T_\mathrm{vir}^\mathrm{min}}=6.7$ corresponds to a minimum halo mass of $~10^{12}M_\odot$.}\par
In Figure \ref{eor_params}, we present the posterior distributions of $\zeta$ and $T_\mathrm{vir}^\mathrm{min}$.
Despite a strong degeneracy between these two parameters, we derive \Add{$\log{\zeta}=2.50_{-0.96}^{+1.61}$ and $\log{T_\mathrm{vir}^\mathrm{min}}/\mathrm{K}=5.61_{-0.05}^{+1.01}$}.
Figure \ref{fit_crh} shows the reionization history derived from the best-fit EoR parameters, which successfully reproduces the measured $\xHI$ values.
\Add{At $z\sim7$, the best-fit scenario lies between the constraints of this study and those from \citet{umeda24b}.}
Additionally, the optical depth of the corresponding reionization scenario, $\tau_e=0.054$, agrees with the Planck result.\par
\Add{Note that our $\xHI$ values are derived through comparison with a simulation using $\zeta=20$ and $T_\mathrm{vir}^\mathrm{min}=5\times10^4~\mathrm{K}$, which are inconsistent with the obtained best-fit parameters.
To address this discrepancy, We run a simulation with the best-fit parameters and re-estimate the $\xHI$ values.
The recalculated $\xHI$ values are $<0.41$, $0.70_{-0.32}^{+0.20}$, $0.86_{-0.18}^{+0.12}$, and $0.89_{-0.13}^{+0.11}$ at $z\sim6$, $7$, $8-9$, and $10-14$, respectively.
These values are consistent with those derived in Section \ref{sec;xhi_estimate} within errors.
Therefore, we confirm that the selection of input parameters does not significantly affect the estimation of $\xHI$ values, and that our $\xHI$ values and the best-fit reionization history are self-consistent.}\par
From Equation (\ref{temp_mass}), the best-fit value of the minimum virial temperature, $T_\mathrm{vir}^\mathrm{min}=10^{5.6}~\mathrm{K}$, corresponds to a minimum halo mass of $M_h^\mathrm{min}=10^{10.5}M_\odot$ at $z=6.5$.
This implies that only halos more massive than $10^{10.5}M_\odot$ are capable of emitting ionizing photons beyond their boundaries.
Using Equation (\ref{magmas}), the UV magnitude of the faintest ionizing sources is estimated to be $M_\mathrm{UV}=-17$.
Since our galaxy sample at $z\sim6.5$ includes galaxies with $M_\mathrm{UV}>-17$, this threshold luminosity is brighter than our current observational limit.
Previous studies often assumed that all star-forming galaxies contribute to ionizing photon escape and that the minimum halo mass for ionizing photon emission is the same as the minimum halo mass required for star formation through efficient cooling.
For instance, \citet{ishigaki18} derive a lower limit for the threshold UV magnitude of $M_\mathrm{UV}>-14.0$ under this assumption.
However, our brighter threshold value of $M_\mathrm{UV}=-17$ suggests that the minimum mass for ionizing sources is larger than \Add{the one} for galaxy-hosting halos. This implies that the faintest galaxies do not significantly contribute to ionizing photon escape.\par
We then consider the implications of $\zeta\sim10^{2.5}$.
In the following discussion, we consider two types of ionizing sources: star-forming galaxies and \Add{active galactic nuclei (AGNs)}.
First, assuming that star-forming galaxies are the dominant sources, we connect $\zeta$ to the ionizing photon production efficiency $\xi_\mathrm{ion}$, which is defined as the ratio of the ionizing photon emission rate to the UV luminosity, and the escape fraction of ionizing photons, $f_\mathrm{esc}$.
By multiplying the number of baryons and differentiating Equation (\ref{zeta}) with respect to the look-back time $t$, we derive the following equation:
\begin{align}\label{zeta_to_xi}
    \zeta\dv{M_h}{t}\frac{\Omega_bX_p}{\Omega_mm_p}=f_\mathrm{esc}\xi_\mathrm{ion}L_\mathrm{UV},
\end{align}
where $M_h$ is the halo mass, $m_p$ is the proton mass, and $L_\mathrm{UV}$ is the UV luminosity of a galaxy residing in a halo \Add{with} mass $M_h$.
Hereafter, we focus on $\xi_\mathrm{ion}$ and $f_\mathrm{esc}$ at $z\sim6.5$, although this method can also be applied to other redshifts.
We adopt the mean halo mass growth rate derived by \citet{fakhouri10}:
\begin{align}
    \expval{\dv{M\Add{_h}}{t}}=&46.1~M_\odot\ \mathrm{yr^{-1}}\qty(\frac{M_h}{10^{12}~M_\odot})^{1.1}\notag\\
    &\times(1+1.11z)\sqrt{\Omega_m(1+z)^3+\Omega_\Lambda}.
\end{align}
Using the halo mass-UV magnitude relation from Equation (\ref{magmas}), both $M_h$ and $L_\mathrm{UV}$ can be expressed as functions of $M_\mathrm{UV}$.
\Add{From observations,} \citet{simmonds24} derive $\xi_\mathrm{ion,0}$, which is the ionizing photon production efficiency under the assumption of an escape fraction of zero, as a function of UV magnitude for $6<z\leq7$:
\begin{align}
    \log{\xi_\mathrm{ion,0}~\mathrm{[Hz\ erg^{-1}]}}=-0.03M_\mathrm{UV}+24.88.
\end{align}
The relationship between $\xi_\mathrm{ion}$ and $\xi_\mathrm{ion,0}$ is given by:
\begin{align}\label{xi_xi0}
    \xi_\mathrm{ion,0}=(1-f_\mathrm{esc})\xi_\mathrm{ion}.
\end{align}
Substituting this into Equation (\ref{zeta_to_xi}) gives the following expression for $f_\mathrm{esc}$:
\begin{align}\label{fesc_equ}
    f_\mathrm{esc}=\frac{1}{1+\frac{\xi_\mathrm{ion,0}L_\mathrm{UV}\Omega_mm_p}{\zeta(\dv*{M_h}{t})\Omega_bX_p}}.
\end{align}
\par
\begin{figure}
    \centering
    \includegraphics[width=\linewidth]{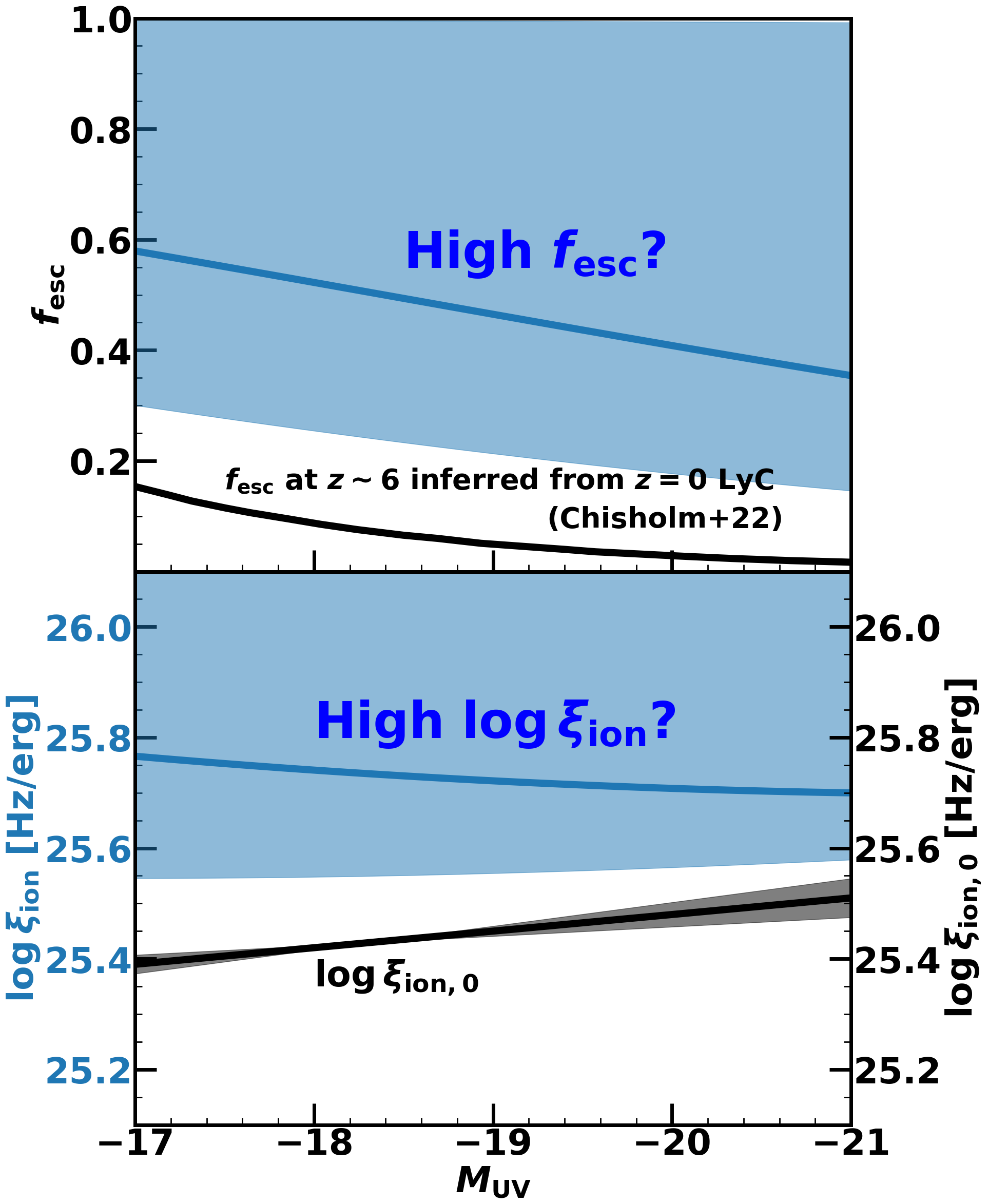}
    \caption{Lyman continuum escape fraction $f_\mathrm{esc}$ (top) and ionizing photon production efficiency $\xi_\mathrm{ion}$ (bottom) of star-forming galaxies at $z\sim6.5$, as a function of UV magnitude. The blue lines indicate $f_\mathrm{esc}$ and $\xi_\mathrm{ion}$ obtained in this work. The blue shaded regions represent the $1\sigma$ errors. \Add{These quantities are derived based on the $\zeta$ values in Figure \ref{eor_params} and the ionizing photon production efficiency assuming an escape fraction of zero, $\xi_\mathrm{ion,0}$, obtained by \citet{simmonds24}. The black line in the bottom panel shows $\xi_\mathrm{ion,0}$ obtained by \citet{simmonds24}. Our $\xi_\mathrm{ion}$ values are higher than $\xi_\mathrm{ion,0}$ because the effect of the high $f_\mathrm{esc}$ is non-negligible in Equation (\ref{xi_xi0}). The black line in the top panel shows $f_\mathrm{esc}$ at $z=6$, which is inferred from the Low-redshift Lyman Continuum Survey (LzLCS) \citep{chisholm22}. The $f_\mathrm{esc}$ estimated from the reionization history in this work and LzLCS in \citet{chisholm22} exhibit a clear tension.}}
    \label{fesc}
\end{figure}

In the top panel of Figure \ref{fesc}, we show $f_\mathrm{esc}$ at $z\sim6.5$ as a function of $M_\mathrm{UV}$.
The blue solid line and the shaded region correspond to the best-fit and $1\sigma$ uncertainty of the posterior distribution of $\zeta$ in Figure \ref{eor_params}.
The large $\zeta$ value obtained in this work implies a high escape fraction of $f_\mathrm{esc}\sim50\%$.
In contrast, \citet{chisholm22} derive the relation between $f_\mathrm{esc}$ and the UV spectral slope $\beta$ \Add{from} the Low-redshift Lyman Continuum Survey (LzLCS) and literature observations at $z\sim0$.
They infer $f_\mathrm{esc}$ at $z\sim6$, shown as the black solid line in the top panel of Figure \ref{fesc}, from the relation.
The $f_\mathrm{esc}$ derived from these local galaxy observations is approximately $10\%$, significantly lower than the value estimated in this work.
\Add{\citet{harikane20} directly derive an upper limit of the escape fraction at $z\sim6$, $f_\mathrm{esc}<0.15\pm0.16$, through interstellar absorption lines in the composite spectrum of $z\sim6$ galaxies with $M_\mathrm{UV}\sim-23$, which is also lower than our estimate.}
\Add{These discrepancies suggest} that the late and sharp reionization history derived in this work cannot be explained by star-forming galaxies without assuming \Add{an implausibly sharp} redshift evolution in galaxy properties.
In the bottom panel of Figure \ref{fesc}, we plot $\xi_\mathrm{ion}$ calculated using $f_\mathrm{esc}$ from the top panel and $\xi_\mathrm{ion,0}$ derived by \citet{simmonds24}.
Our $\xi_\mathrm{ion}$ values are $\sim0.3$ dex higher than $\xi_\mathrm{ion,0}$ because the effect of the high $f_\mathrm{esc}$ is non-negligible in Equation (\ref{xi_xi0}).
\par
\Add{We compare our results to one of the most sophisticated and precise reionization simulations, \texttt{THESAN}.
The \texttt{THESAN} simulations \citep{kannan22, garaldi22, smith22} include several runs, and our $\xHI$ values align with the sharpest scenario: THESAN-HIGH-2.
In this simulation, only massive halos ($>10^{10}M_\odot$) contribute to reionization, and their escape fraction is $f_\mathrm{esc}=0.80$.
Therefore, the high $T_\mathrm{vir}^\mathrm{min}$ and $f_\mathrm{esc}$ values derived with \texttt{21cmFAST} above are consistent with the \texttt{THESAN} simulations.
Explaining our late and sharp reionization remains challenging unless an extremely high $f_\mathrm{esc}$ is assumed.}\par
\Add{We note that recent observations report that the cosmic star formation rate at $z\gtrsim10$ is higher than model predictions assuming constant star formation efficiencies (e.g., \citealt{harikane24}).
This suggests that the number of bright galaxies is larger than predicted by Equation (\ref{magmas}) used in this work, highlighting the challenge of reproducing our late reionization history.}\par
Second, we assume that AGNs are the primary reionization sources, rather than star-forming galaxies.
In this case, the time derivative of Equation (\ref{zeta}) is expressed as:
\begin{align}\label{agn_equ}
    \zeta\dv{M_h}{t}\frac{\Omega_bX_p}{\Omega_mm_p}=f_\mathrm{AGN}f_\mathrm{esc}\dot{n},
\end{align}
where $f_\mathrm{AGN}$ represents the fraction of halos with mass $M_h$ hosting AGNs at redshift $z$, and $\dot{n}$ is the ionizing photon emission rate of AGNs.
\Add{A high $f_\mathrm{AGN}$ value implies a large AGN duty cycle.
In the following discussion, we connect $f_\mathrm{AGN}$ to the best-fit $\zeta$ value.}
To link halo mass with AGN UV magnitude,
we use the AGN UV flux ratio $f_\mathrm{UV,AGN}$ obtained by \citet{zhang21}, which quantifies the contribution of AGN UV flux to the total UV flux (including the host galaxy):
\begin{align}
    \log{f_\mathrm{UV,AGN}}=-0.118M_\mathrm{UV}-3.056,
\end{align}
where $M_\mathrm{UV}$ is the AGN+host galaxy UV magnitude.
Using this relation, the AGN UV luminosity can be determined as a function of the host galaxy's UV luminosity.
Assuming that the host galaxy's UV luminosity follows Equation (\ref{magmas}), we derive the AGN UV luminosity as a function of halo mass.
We model the \Add{spectral energy distribution (SED)} of AGN as a power law ($f_\nu\propto\nu^\alpha$) with an index of $\alpha=-0.61$ for $\lambda>912$~\AA\ and $\alpha=-1.7$ for $\lambda\leq912$~\AA\ \citep{lusso15}.
This allows us to calculate $\dot{n}$ for a given AGN UV luminosity.\par
\begin{figure}
    \centering
    \includegraphics[width=\linewidth]{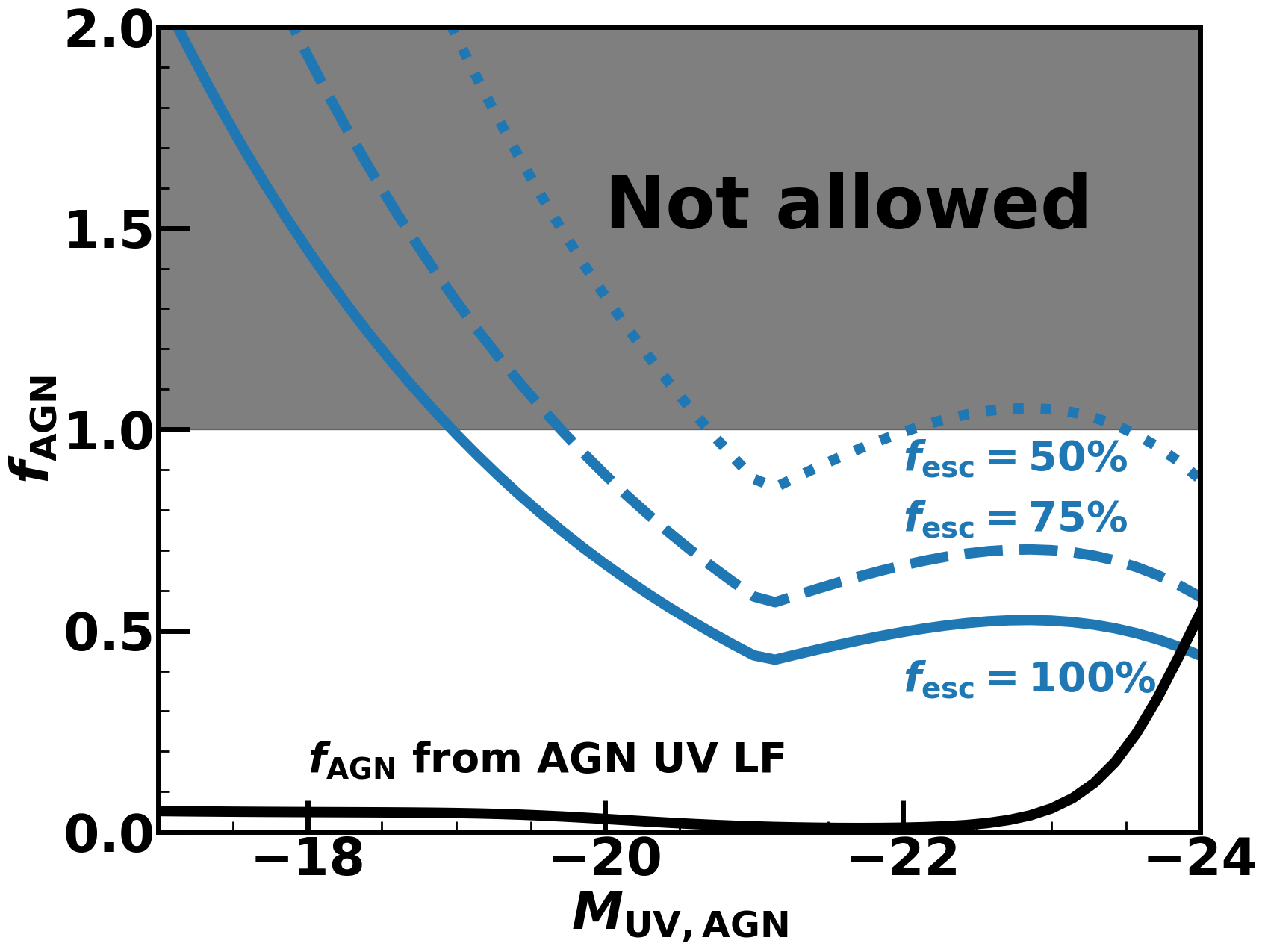}
    \caption{AGN fraction $f_\mathrm{AGN}$ as a function of AGN UV magnitude without the contribution from the host galaxy, at $z\sim6.5$. $f_\mathrm{AGN}$ represents the fraction of halos \Add{with} mass $M_h$ hosting AGNs. The blue \Add{dotted, dashed, and solid} lines correspond to $f_\mathrm{AGN}$ values derived in this work assuming $\zeta=10^{2.5}$ and escape fractions $f_\mathrm{esc}=50\%$, $75\%$, and $100\%$\Add{, respectively}. The gray shaded region indicates unphysical values where $f_\mathrm{AGN}>1$. The black solid line shows $f_\mathrm{AGN}$ values \Add{derived} from the ratio of the galaxy and AGN UV luminosity functions at $z\sim6$. Across most of the UV magnitude range, our $f_\mathrm{AGN}$ values significantly exceed those inferred from luminosity functions.}
    \label{fagn}
\end{figure}

We estimate $f_\mathrm{AGN}$ for three cases: $f_\mathrm{esc}=50\%$, $75\%$, and $100\%$ (Figure \ref{fagn}).
Even with an escape fraction of $100\%$, the estimated $f_\mathrm{AGN}$ at $M_\mathrm{UV}\gtrsim-19$ exceeds $100\%$, indicating that faint AGNs are unlikely to contribute ionizing photons at the efficiency of $\zeta\sim10^{2.5}$.
For comparison, we also calculate $f_\mathrm{AGN}$ using the ratio of the galaxy UV luminosity function \citep{bouwens21} to the AGN UV luminosity function at $z\sim6$. \Add{The black solid line in Figure \ref{fagn} shows $f_\mathrm{AGN}$ derived from the galaxy and AGN UV luminosity functions}.
The AGN luminosity function \Add{at $z\sim6$} combines bright AGNs \citep{matsuoka18} and faint AGNs \citep{harikane23}.
Our $f_\mathrm{AGN}$ values, derived under the assumption of high ionizing efficiency ($\zeta\sim10^{2.5}$), are significantly higher than those derived from luminosity functions.
\Add{Note that we derive $f_\mathrm{AGN}$ assuming $\zeta$ as a constant parameter. If $\zeta$ is larger for brighter AGNs, the problem that $f_\mathrm{AGN}$ exceeds $100\%$ for faint AGNs may be solved. However, Solving the tension with $f_\mathrm{AGN}$ derived from luminosity functions is difficult because our $f_\mathrm{AGN}$ values are higher than those derived from luminosity functions across most of the UV magnitude range.}

\subsection{Possible Explanations}\label{sec;sum_dis}
In Section \ref{sec;recombination} and \ref{sec;massivehalos}, we demonstrate that neither star-forming galaxies in massive halos, AGNs in massive halos, nor strong recombination cannot fully account for the late and sharp reionization history derived in this work.
In this subsection, we discuss several potential solutions to address the discrepancy.\par
First, the escape fraction of star-forming galaxies during the EoR may be indeed as high as $f_\mathrm{esc}\sim50\%$, as suggested in Figure \ref{fesc}.
Although this value is inconsistent with predictions based on observations of local galaxies \citep{chisholm22}, galaxy properties can evolve with redshift.
For instance, \citet{naidu22} show that approximately half of LAEs at $z\sim2$ exhibit $f_\mathrm{esc}\sim50\%$. This may result from ionizing photons produced by massive and young stars escaping feedback-induced holes in the ISM.
While such LAEs constitute only a small fraction of galaxies at $z\sim2$, the late and sharp reionization history could be explained if these galaxies dominate the universe during the EoR.
See also \citet{naidu20} and \citet{matthee22} for reionization models dominated by bright galaxies or LAEs.\par
Second, the duty cycle of faint AGNs may be significantly larger than currently understood.
Although JWST observations have identified numerous faint broad-line AGNs, the number density could be even larger when narrow-line type 2 AGNs are included (e.g., \citealt{scholtz23}).
Ionizing photons emitted by these faint type 2 AGNs could address the reionization challenge.
\Add{\citet{asthana24} and \citet{madau24} also propose late and sharp reionization scenarios where faint AGNs dominate (or partially contribute to) the photon budget, although our $\zeta$ value exceeds theirs.}\par
\Add{Third, $f_\mathrm{esc}$ might evolve with redshift. If $f_\mathrm{esc}$ increases toward lower redshift, the late and sharp reionization can be explained.
For example, \citet{yajima14} investigate the redshift evolution of $f_\mathrm{esc}$ with simulations and show a mild increase from $z\sim10$ to $z\sim7$, attributed to the lower gas density in galaxies at lower redshift.
Another possibility is that external UV radiation ionizes the outer layers of dwarf galaxies at lower redshift, potentially resulting in a higher $f_\mathrm{esc}$.}\par
\Add{Fourth, our $\xHI$ values may be biased.
To derive models of EW distributions, we assume that the peak velocity of Ly$\alpha$ emission follows a log-normal distribution, with its peak given by Equation (\ref{vel_cen}).
This probability distribution depends solely on $M_h$.
However, it is possible that the peak velocity also evolves with redshift.
The velocity offset of Ly$\alpha$ emission originates from interactions with galactic outflows.
Therefore, the offset could be smaller at higher redshifts if outflows are weaker.
\revise{Observationally, \citet{vitte25} report that the peak separation between the blue and red components of Ly$\alpha$ emission decreases with redshift, although \citet{hayes21} do not find a significant redshift evolution in the Ly$\alpha$ line profile.
If the velocity offset indeed decreases with redshift,} Ly$\alpha$ photons would experience stronger absorption at higher redshifts.
Consequently, the observed evolution of Ly$\alpha$ emission could be explained without adopting high $\xHI$ values.
}\par
\Add{Lastly, alternative sources beyond star-forming galaxies and AGNs may significantly contribute to reionization and produce a late and sharp reionization history.
Potential contributors include globular clusters with $f_\mathrm{esc}\sim1$ (e.g., \citealt{ricotti02}), high-mass X-ray binaries (e.g., \citealt{jeon14}), and primordial black holes (e.g., \citealt{ricotti08, tashiro13, clark18}).
The late and sharp reionization could be explained if these sources efficiently produce ionizing photons at $z\sim6-8$.}

\section{Summary}\label{sec;summary}
In this paper, we present our constraints on $\xHI$ from Ly$\alpha$ EWs of galaxies at $z\sim5-14$.
We select \Add{586} galaxies at $z=4.5-14.2$ observed by multiple JWST/NIRSpec spectroscopy projects, JADES, GLASS, CEERS, and GO/DDT programs.
The redshifts of galaxies in our sample are determined with emission lines and/or a strong continuum break.
We fit the galaxy spectra near rest-frame 1216~\AA\ to obtain Ly$\alpha$ EW measurements or upper limits for all these galaxies.
Our major findings are summarized below:
\begin{itemize}
    \item[1.] We find that the fraction of LAEs with $\mathrm{EW}>25$~\AA\ (10~\AA) decreases from $22_{-4}^{+5}\%$ ($29_{-4}^{+7}\%$) at $z\sim5$ to $<10\%$ ($10_{-8}^{+13}\%$) at $z\sim10-14$ for $-20.25<M_\mathrm{UV}<-18.75$ galaxies, suggesting increasing Ly$\alpha$ damping wing absorption at higher redshift.
    The Ly$\alpha$ fractions obtained in this work are consistent with past JWST measurements.
    \Add{The decreasing trend of Ly$\alpha$ fraction is more clearly seen for fainter ($-18.75<M_\mathrm{UV}<-17.25$) galaxies, likely due to the smaller ionized bubble radius around faint galaxies.}
    \item[2.] We derive the Ly$\alpha$ escape fraction $f_\mathrm{esc}^\mathrm{Ly\alpha}$.
    We find that $f_\mathrm{esc}^\mathrm{Ly\alpha}$ decreases with redshift and increases with Ly$\alpha$ EW, suggesting increasing Ly$\alpha$ damping wing absorption at higher redshift.
    We also find $f_\mathrm{esc}^\mathrm{Ly\alpha}$ values are low for galaxies with a red UV spectrum ($\beta$) and high ionizing photon production efficiency assuming zero LyC escape ($\xi_\mathrm{ion,0}$), which can be explained by a thick gaseous disc.
    \item[3.] We derive Ly$\alpha$ luminosity functions at $z=5-14$ with the observed Ly$\alpha$ EW distributions and galaxy UV luminosity functions.
    \revise{We present the first constraints on the Ly$\alpha$ luminosity function at $z\sim8-14$ and find that} the Ly$\alpha$ luminosity function at $L_\mathrm{Ly\alpha}=10^{42}-10^{43}~\mathrm{erg\ s^{-1}}$ decreases by $\sim3$ dex from $z\sim5$ to $z\sim10-14$. 
    \item[4.] To obtain EW distribution models during the EoR, We use the EW distribution at $z\sim5$ as the intrinsic Ly$\alpha$ EW distribution and model Ly$\alpha$ absorption in the IGM at $z\gtrsim6$ \Add{calculated} with \texttt{21cmFAST} semi-numerical simulations that reproduce the observed UV luminosity function during the EoR and the CMB optical depth.
    We obtain $\xHI=0.17_{-0.16}^{+0.23}$, $0.63_{-0.28}^{+0.18}$, $0.79_{-0.21}^{+0.13}$, and $0.88_{-0.13}^{+0.11}$ at $z\sim6$, $7$, $8-9$, and $10-14$, respectively, via the comparisons of the observed EW distributions with the models.
    Our high $\xHI$ values at $z\sim7-14$ suggest a very late and sharp reionization.
    \item[5.] Assuming star-forming galaxies or AGNs as major ionizing sources, the late and sharp reionization history obtained in this study \Add{is} explained with massive halos ($M_h>10^{10.5}M_\odot$) as the main ionizing sources.
    \Add{However, an extremely} high escape fraction $f_\mathrm{esc}\sim50\%$ and ionizing photon production efficiency (large duty cycle) are required for galaxies (AGNs) \Add{at the same time}.
    We also confirm that \Add{a high recombination rate} cannot explain our $\xHI$ measurements.
    Therefore\Add{,} a redshift evolution of the escape fraction, a high number density of obscured AGNs, or alternative ionizing sources may be needed.
\end{itemize}

%% Also note that the akcnowlodgment environment does not support long amounts of text. If you have a lot of people and institutions to acknowledge, do not use this command. Instead, create a new \section{Acknowledgments}.
\section*{Acknowledgements}
\revise{We thank the anonymous referee for the valuable comments that greatly improved this manuscript.}
We thank Sarah Bosman, Akio Inoue, Laura Keating, Charlotte Mason, Kana Moriwaki, Hyunbae Park, Hayato Shimabukuro, and Daniel Stark for valuable discussions on this work. \Add{We thank Yuki Isobe for providing us with the table of the line spread function of the NIRpec instrument. We thank Hiroto Yanagisawa for providing us with the table of the UV slope measurements.} This work is based on observations made with the NASA/ESA/CSA James Webb Space Telescope. The data were obtained from the Mikulski Archive for Space Telescopes at the Space Telescope Science Institute, which is operated by the Association of Universities for Research in Astronomy, Inc., under NASA contract NAS 5-03127 for JWST. These observations are associated with programs GTO 1180/1181 (JADES; PI: D. Eisenstein), GTO 1210/1286 (JADES; PI: N. Lützgendorf), GO 3215 (JADES; PI: D. Eisenstein \& R. Maiolino), ERS 1324 (GLASS; PI: T. Treu), ERS 1345 (CEERS; PI: S. Finkelstein), GO 1433 (PI: D. Coe), and DDT 2750 (PI: P. Arrabal Haro). The authors acknowledge the teams conducting these observations for developing their observation programs. We thank the JADES team for publicly releasing reduced spectra and catalogs from the JADES survey. This publication is based on work supported by the World Premier International Research Center Initiative (WPI Initiative), MEXT, Japan, KAKENHI (\Add{21H04489, 24H00245, 24K00625, 24K17098}) through the Japan Society for the Promotion of Science, and JST FOREST Program (JP-MJFR202Z). This work was supported by the joint research program of the Institute for Cosmic Ray Research (ICRR), the University of Tokyo. This research was supported by FoPM, WINGS Program, the University of Tokyo.
\software{21cmFAST \citep{mesinger11, murray20}, Astropy \citep{astropy13, astropy18, astropy22}, corner \citep{foremanmackey16}, emcee \citep{foremanmackey13}, matplotlib \citep{hunter07}, NumPy \citep{harris20}, \Add{PyNeb \citep{luridiana15}}, scikit-learn \citep{pedregosa11}, and SciPy \citep{virtanen20}.}

\appendix
\restartappendixnumbering
\section{Sample in this study}
\begin{figure*}[t]
    \centering
    \includegraphics[width=\linewidth]{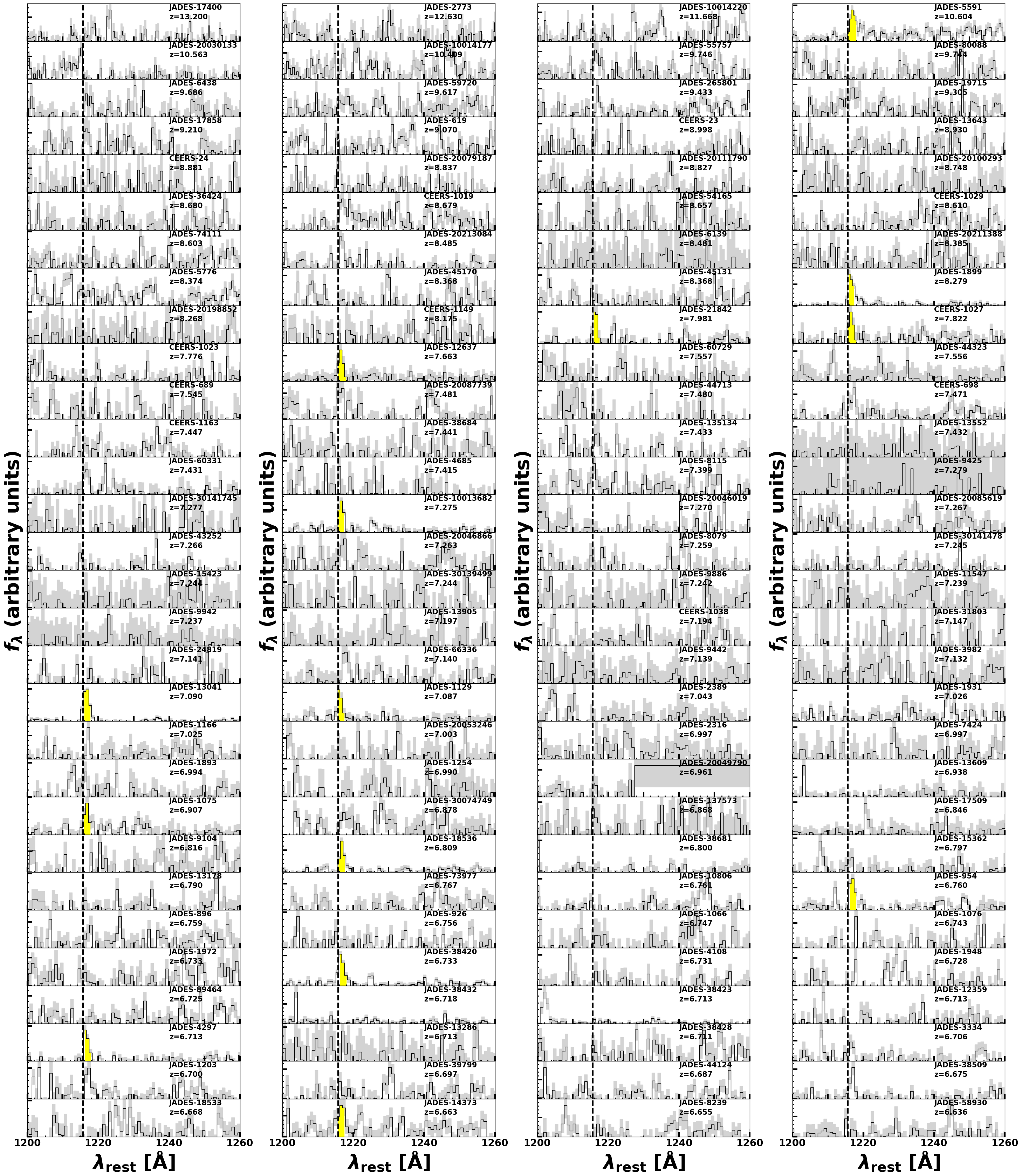}
    \caption{Spectra of the galaxies in our sample. The black solid lines and shaded regions represent the observed spectra and associated $1\sigma$ errors, respectively. The vertical dashed lines represent the rest-frame Ly$\alpha$ wavelength. The yellow regions indicate the detected Ly$\alpha$ lines.}
    \label{spec1}
\end{figure*}
\setcounter{figure}{0}
\begin{figure*}[t]
    \centering
    \includegraphics[width=\linewidth]{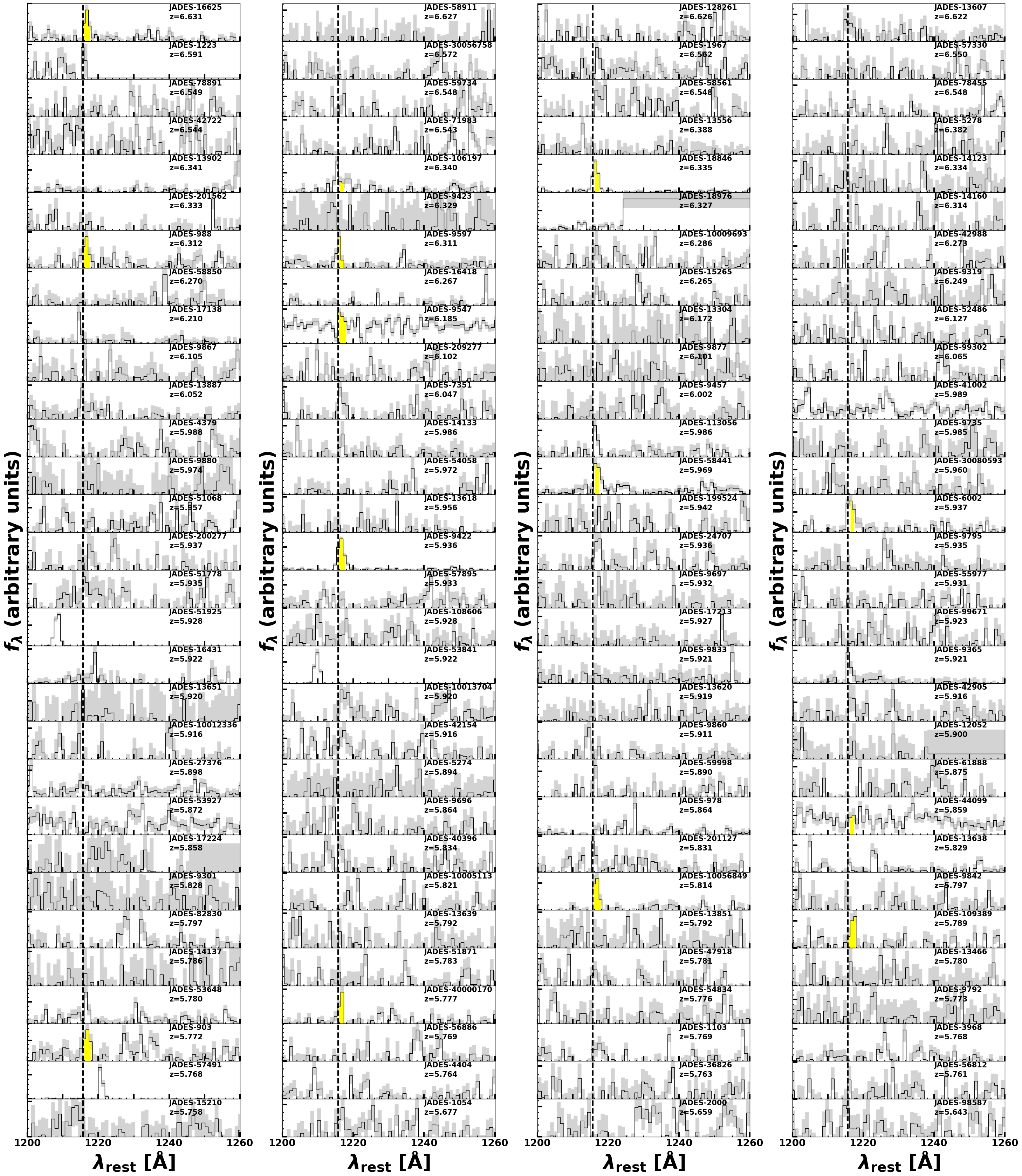}
    \caption{Continued.}
    \label{spec2}
\end{figure*}
\setcounter{figure}{0}
\begin{figure*}[t]
    \centering
    \includegraphics[width=\linewidth]{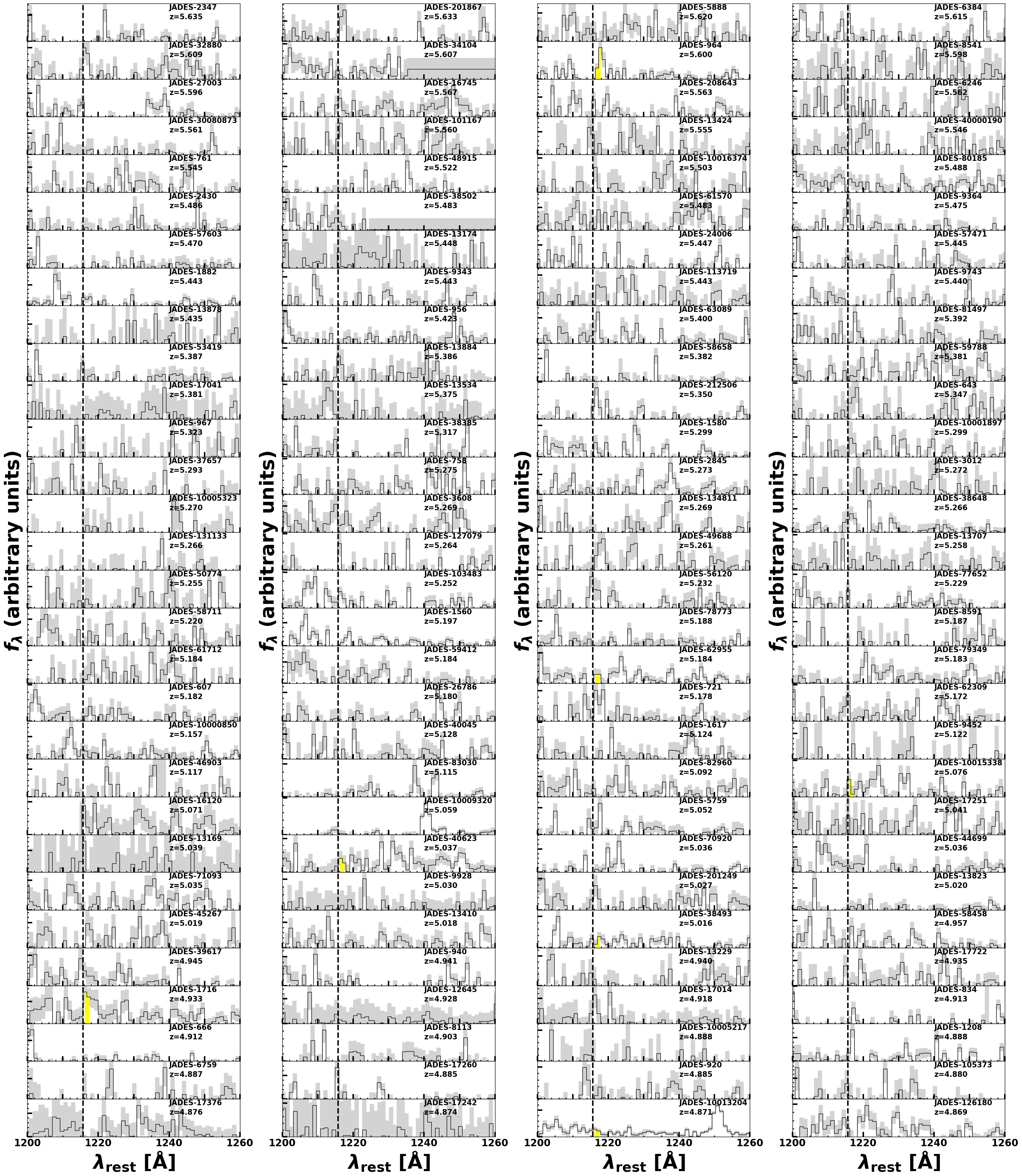}
    \caption{Continued.}
    \label{spec3}
\end{figure*}
\setcounter{figure}{0}
\begin{figure*}[t]
    \centering
    \includegraphics[width=\linewidth]{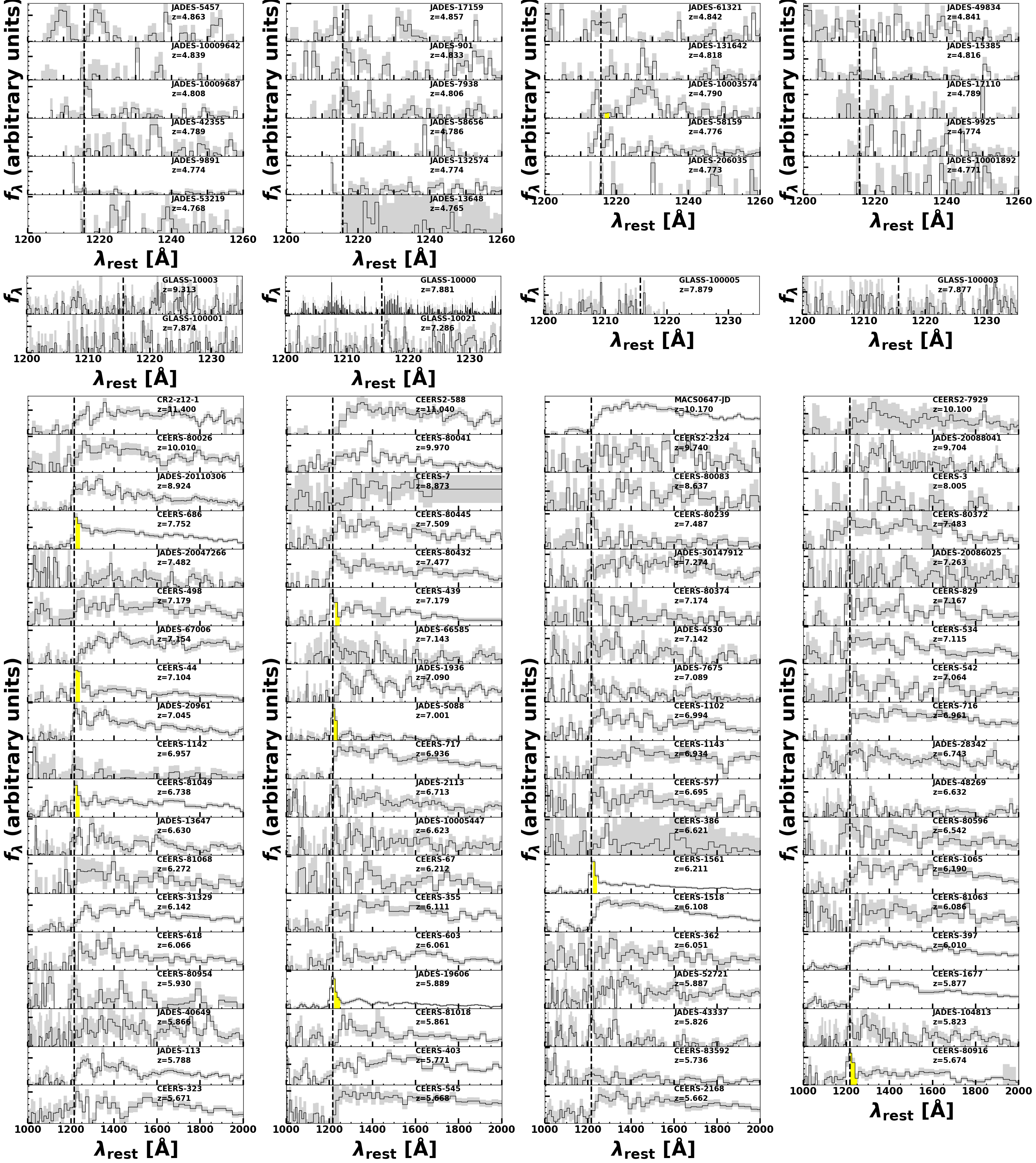}
    \caption{Continued.}
    \label{spec4}
\end{figure*}
\setcounter{figure}{0}
\begin{figure*}[t]
    \centering
    \includegraphics[width=\linewidth]{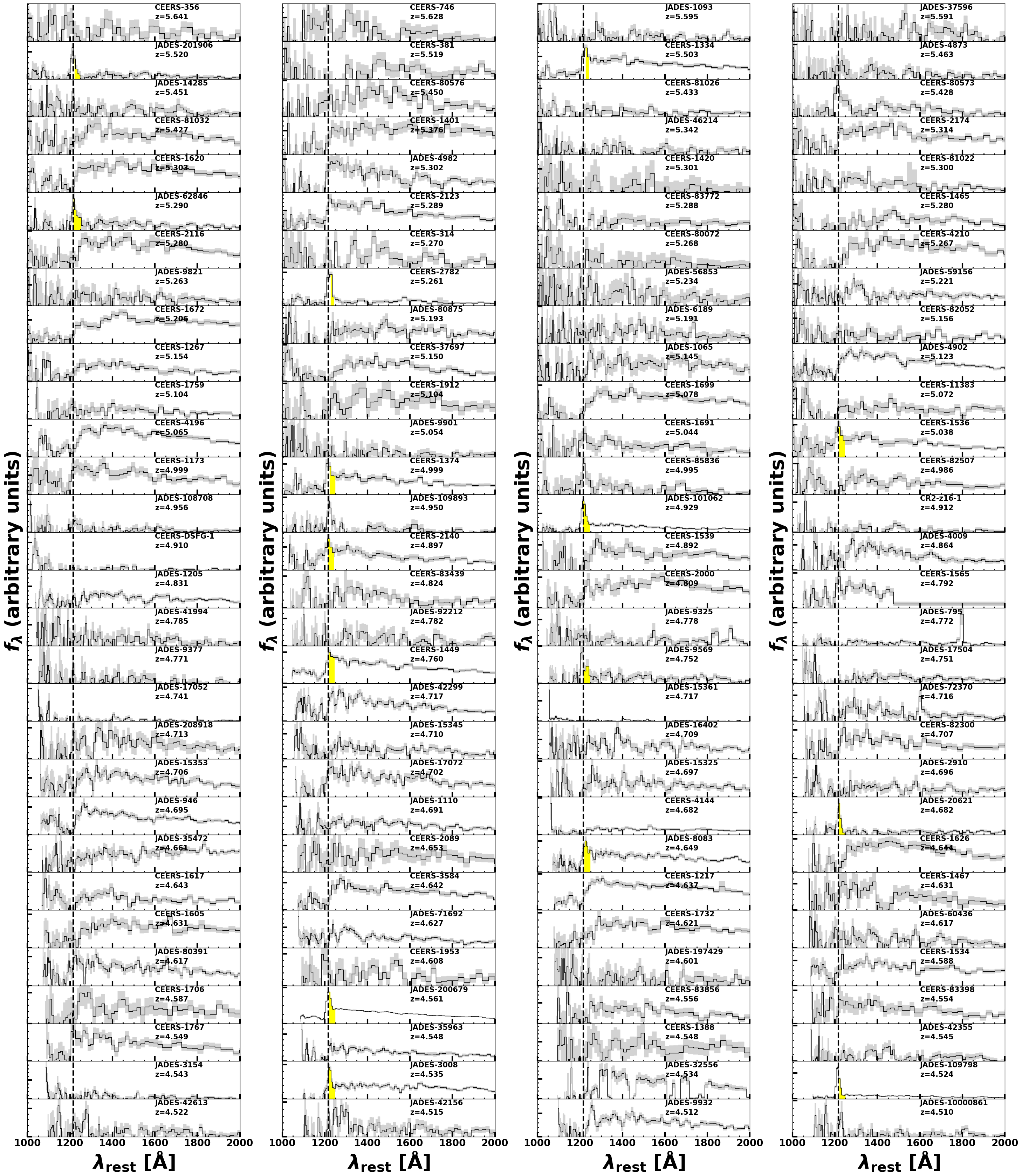}
    \caption{Continued.}
    \label{spec5}
\end{figure*}
%\startlongtable
%\centerwidetable
\begin{longrotatetable}
\begin{deluxetable*}{cccccccccccc}
\tablecaption{Sample in this study}
\tabletypesize{\scriptsize}
\tablehead{ID & R.A. & Decl. & $z_\mathrm{sys}$ & $M_\mathrm{UV}$ & $F_\mathrm{Ly\alpha}$ & $\Delta v_\mathrm{Ly\alpha}$ & $f_\mathrm{esc}^\mathrm{Ly\alpha}$ & $\mathrm{EW_{Ly\alpha}}$ & $\mathrm{EW_{Ly\alpha}}$ & $\mathrm{EW_{Ly\alpha}}$ & PID\\
 & (deg) & (deg) & & (mag) & ($10^{-19}\mathrm{erg\ s^{-1}\ cm^{-2}}$) & ($\mathrm{km\ s^{-1}}$) & & (\AA, This work) & (\AA, Tang24) & (\AA, Jones24) & \\
 (1) & (2) & (3) & (4) & (5) & (6) & (7) & (8) & (9) & (10) & (11) & (12)
}
\startdata
JADES-GS-z14-0 & $53.08294$ & $-27.85563$ & $14.180^\mathrm{a}$ & $-20.81^\mathrm{b}$ & & & & $<5^\mathrm{b}$ & & & 1287\\
JADES-GS-z14-1 & $53.07427$ & $-27.88592$ & $13.90^\mathrm{b}$ & $-19.0^\mathrm{b}$ & & & & $<10^\mathrm{b}$ & & & 1287\\
JADES-17400 & $53.14988$ & $-27.77650$ & $13.200$ & $-18.99$ & $<10$ & & & $<16$ & $<45$ & & 1210\\
JADES-GS-z13-1-LA & $53.06475$ & $-27.89024$ & $13.01^\mathrm{c}$ & $-18.66^\mathrm{c}$ & & $696\pm299^\mathrm{c}$ & $0.06\pm0.02^\mathrm{c}$ & $43\pm13^\mathrm{c}$ & & $<242$ & 1287\\
JADES-2773 & $53.16634$ & $-27.82156$ & $12.630$ & $-18.14$ & $<2$ & & & $<11$ & $<103$ & & 1210\\
JADES-10014220 & $53.16477$ & $-27.77463$ & $11.668$ & $-19.30$ & $<4$ & & & $<24$ & $<45$& & 1210\\
CR2-z12-1 & $214.94317$ & $52.94244$ & $11.400$ & $-19.76$ & $<4$ & & & $<8$ & $<26$ & & 2750\\
CEERS2-588 & $214.90662$ & $52.94550$ & $11.040$ & $-19.77$ & $<3$ & & & $<4$ & $<38$ & & 2750\\
\enddata
\tabletypesize{\small}
\tablecomments{(1) Galaxy ID. (2) R.A. of the object’s coordinate. (3) Decl. of the object’s coordinate. (4) Systemic redshift. (5) Absolute UV magnitude. (6) Ly$\alpha$ flux and the $1\sigma$ error. For galaxies with no Ly$\alpha$ detections, we show $3\sigma$ upper limits. (7) Ly$\alpha$ velocity offset. (8) Ly$\alpha$ escape fraction. (9) Rest-frame Ly$\alpha$ EW measured in this work. EWs of JADES-GS-z14-0, JADES-GS-z14-1, JADES-GS-z13-1-LA measured by \citet{carniani24a} and \citet{witstok24} are included in this column. (10) Rest-frame Ly$\alpha$ EW measured by \citet{tang24b}. (11) Rest-frame Ly$\alpha$ EW measured by \citet{jones24}. (12) JWST program ID.\\
The values of UV magnitude, Ly$\alpha$ flux, and EW are not corrected for slit loss.\\
a: \citet{carniani24b}, b: \citet{carniani24a}, c: \citet{witstok24}, d: \citet{harikane24}, e: \citet{nakajima23}}
\label{tab:sample}
\end{deluxetable*}
\end{longrotatetable}

\clearpage




\bibliography{kageura25}{}
\bibliographystyle{aasjournal}



\end{document}